\documentclass[aip,reprint,jcp,numerical]{revtex4-1}
\makeatletter\if@twocolumn\PassOptionsToPackage{switch}{lineno}\else\fi\makeatother

\usepackage{tabulary,graphicx,amsfonts,amsmath,amssymb,amsbsy}
\usepackage[T1]{fontenc}

\usepackage{url,multirow,morefloats,floatflt,cancel,tfrupee}
\makeatletter

\AtBeginDocument{\@ifpackageloaded{textcomp}{}{\usepackage{textcomp}}}
\makeatother
\usepackage{colortbl}
\usepackage{xcolor}
\usepackage{pifont}
\usepackage[nointegrals]{wasysym}
\makeatletter

\def\mcWidth#1{\csname TY@F#1\endcsname+\tabcolsep}

\def\cAlignHack{\rightskip\@flushglue\leftskip\@flushglue\parindent\z@\parfillskip\z@skip}
\def\rAlignHack{\rightskip\z@skip\leftskip\@flushglue \parindent\z@\parfillskip\z@skip}

\@ifundefined{etal}{}{}

\usepackage{ifxetex}
\ifxetex\else\if@twocolumn\@ifpackageloaded{stfloats}{}{\usepackage{dblfloatfix}}\fi\fi

\AtBeginDocument{
\expandafter\ifx\csname eqalign\endcsname\relax
\def\eqalign#1{\null\vcenter{\def\\{\cr}\openup\jot\m@th
  \ialign{\strut$\displaystyle{##}$\hfil&$\displaystyle{{}##}$\hfil
      \crcr#1\crcr}}\,}
\fi
}

\AtBeginDocument{%
  \@ifpackageloaded{endfloat}%
   {\renewcommand\efloat@iwrite[1]{\immediate\expandafter\protected@write\csname efloat@post#1\endcsname{}}}{\newif\ifefloat@tables}%
}%

\def\BreakURLText#1{\@tfor\brk@tempa:=#1\do{\brk@tempa\hskip0pt}}
\let\lt=<
\let\gt=>
\def\processVert{\ifmmode|\else\textbar\fi}

\@ifundefined{subparagraph}{
\def\subparagraph{\@startsection{paragraph}{5}{2\parindent}{0ex plus 0.1ex minus 0.1ex}%
{0ex}{\normalfont\small\itshape}}%
}{}

\newcommand\role[1]{\unskip}
\newcommand\aucollab[1]{\unskip}
  
\@ifundefined{tsGraphicsScaleX}{\gdef\tsGraphicsScaleX{1}}{}
\@ifundefined{tsGraphicsScaleY}{\gdef\tsGraphicsScaleY{.9}}{}
\def\checkGraphicsWidth{\ifdim\Gin@nat@width>\linewidth
	\tsGraphicsScaleX\linewidth\else\Gin@nat@width\fi}

\def\checkGraphicsHeight{\ifdim\Gin@nat@height>.9\textheight
	\tsGraphicsScaleY\textheight\else\Gin@nat@height\fi}

\def\fixFloatSize#1{}
\let\ts@includegraphics\includegraphics

\def\inlinegraphic[#1]#2{{\edef\@tempa{#1}\edef\baseline@shift{\ifx\@tempa\@empty0\else#1\fi}\edef\tempZ{\the\numexpr(\numexpr(\baseline@shift*\f@size/100))}\protect\raisebox{\tempZ pt}{\ts@includegraphics{#2}}}}

\AtBeginDocument{\def\includegraphics{\@ifnextchar[{\ts@includegraphics}{\ts@includegraphics[width=\checkGraphicsWidth,height=\checkGraphicsHeight,keepaspectratio]}}}

\DeclareMathAlphabet{\mathpzc}{OT1}{pzc}{m}{it}

\def\URL#1#2{\@ifundefined{href}{#2}{\href{#1}{#2}}}

\def\UrlOrds{\do\*\do\-\do\~\do\'\do\"\do\-}%
\g@addto@macro{\UrlBreaks}{\UrlOrds}

\edef\fntEncoding{\f@encoding}

\makeatother

\def\style#1#2{#2}

\newif\ifmultipleabstract\multipleabstractfalse%
\usepackage{float,newfloat}

\newcommand{\texttildeapprox}{{\fontfamily{pcr}\selectfont\texttildelow}}

\begin{document}

\def\authorCount{3}
\def\affCount{1}

\def\journalTitle{The Journal of Chemical Physics}

\title{\textbf{Propagation of maximally localized Wannier functions in real-time TDDFT}}
\author{Dillon C. Yost}
\affiliation{Department of Chemistry\unskip, The University of North Carolina at Chapel Hill\unskip, Chapel Hill\unskip, 27599\unskip, NC\unskip, USA}
\author{Yi Yao}
\affiliation{Department of Chemistry\unskip, The University of North Carolina at Chapel Hill\unskip, Chapel Hill\unskip, 27599\unskip, NC\unskip, USA}
\author{Yosuke Kanai}
\email{ykanai@unc.edu}
\affiliation{Department of Chemistry\unskip, The University of North Carolina at Chapel Hill\unskip, Chapel Hill\unskip, 27599\unskip, NC\unskip, USA}

\begin{abstract}
 Real-time, time-dependent density functional theory (RT-TDDFT) has gained popularity as a first-principles approach to study a variety of excited-state phenomena such as optical excitations and electronic stopping. Within RT-TDDFT simulations, the gauge freedom of the time-dependent electronic orbitals can be exploited for numerical and scientific convenience while the unitary transformation does not alter physical properties calculated from the quantum dynamics of electrons. Exploiting this gauge freedom, we demonstrate propagation of maximally localized Wannier functions within RT-TDDFT. We illustrate its great utility through a number of examples including its application to optical excitation in extended systems using the so-called length gauge, interpreting electronic stopping excitation, and simulating electric field-driven quantized charge transport. We implemented the approach within our plane-wave pseudopotential RT-TDDFT module of the QB@LL code, and performance of the implementation is also discussed. 
\end{abstract}\def\keywordstitle{Keywords}

\maketitle 
    
\section{Introduction}
Density functional theory (DFT), based on the Hohenberg-Kohn theorem \unskip~\cite{364506:8093909}, is perhaps the most widely used approach to calculate properties of matter from first principles.  However, DFT is limited to ground state properties, and many of the physical, chemical, and biological processes of interest in modern science and technology involve excited state dynamics.  Such processes include photo-excitation in dye sensitized solar cells \unskip~\cite{364506:8303646}, hot carrier dynamics in nanomaterials \unskip~\cite{364506:8303644}, and ionizing radiation in biological materials \unskip~\cite{364506:8303645}.  While these phenomena can be studied via advanced spectroscopic methods, first-principles simulations based on time-dependent density functional theory (TDDFT) can be used to provide predictive and detailed insights on the atomistic level.  

Currently, time-dependent density functional theory, based on the Runge-Gross theorem \unskip~\cite{364506:8296666}, is one of the most effective and efficient methods for first-principles calculations of excited states and their dynamics \unskip~\cite{364506:8296667,364506:9893093}.  Most often, it is the linear response formulation of TDDFT (LR-TDDFT) \unskip~\cite{364506:8296709} that is used, due to its utility in calculating low-energy excitations of molecules and materials, allowing for the prediction and interpretation of electronic excitations and absorption spectra \unskip~\cite{364506:8297377,364506:8297378,364506:8297379}. However, as its name suggests, linear response TDDFT can only be applied in the linear response (i.e., weak-field) regime and cannot accurately describe processes involving strong fields, such as laser pulses\unskip~\cite{364506:8296799}, and other non-linear response processes \unskip~\cite{364506:8296841,364506:8296924,364506:8296966,364506:8217261}.  Additionally, even in the linear response regime, LR-TDDFT calculations can come at a prohibitive computational expense if the system includes a large number of electrons\unskip~\cite{364506:8297246}, or if a broadband spectrum is desired. This is due to the fact that the calculation must be carried out iteratively in a space of occupied and virtual state dimensions\unskip~\cite{364506:8297122}.  It should be noted, however, that there has been progress in efficient calculations of broadband absorption using the Liouville-Lanczos method \unskip~\cite{364506:9737965}, and there has been some success in using energy-specific TDDFT to calculate high-energy excited states \unskip~\cite{364506:9894709}, even though these methods are still limited to the linear response regime.  

The real-time propagation approach to time-dependent density functional theory (RT-TDDFT) provides an alternative to LR-TDDFT.  Since some of the first uses of real-time propagation approaches in the 1990s, RT-TDDFT has gained popularity for a variety of reasons \unskip~\cite{364506:9894585}. In principle, RT-TDDFT can be used to describe both linear and nonlinear responses of matter to perturbations of any strength. Also, for large systems and certain properties of interest, the RT-TDDFT approach can be more computationally efficient than LR-TDDFT.  A single RT-TDDFT simulation can be used to obtain the entire broad-band absorption spectrum \unskip~\cite{364506:8303643,364506:8093509}, even including core excitations \unskip~\cite{364506:8217170,364506:8217423,364506:8092425}.  Additionally, RT-TDDFT simulations give access to the time-dependent electron density, allowing for molecular-level analysis of the excitation dynamics of interest \unskip~\cite{364506:8297331,364506:8297332,364506:8297380}. Due to these factors, in recent years there has been a surge in applications of RT-TDDFT to a wide range of excited state phenomena such as electronic stopping\unskip~\cite{364506:8217172,364506:8217173,364506:8217215,364506:8217216,364506:8217217,364506:8217218,364506:8217219,364506:8217261}, optical absorption \unskip~\cite{364506:8217306,364506:8303643,364506:8775247,364506:8775289}, core electron excitations \unskip~\cite{364506:8217170,364506:8092425,364506:8217171,364506:8351790}, electronic circular dichroism spectra \unskip~\cite{364506:8297333}, exciton dynamics in nanostructures\unskip~\cite{364506:9894710}, atom-cluster collisions \unskip~\cite{364506:9918743,364506:9918744}, and laser-induced water splitting \unskip~\cite{364506:8217166}.  The promulgation of RT-TDDFT as a means for simulating excited state phenomena has led to its implementation in a variety of electronic structure codes. These include NWChem \unskip~\cite{364506:8217306}, SIESTA\unskip~\cite{364506:8298262}, CP2K\unskip~\cite{364506:8298304}, SALMON \unskip~\cite{364506:8298307}, Octopus \unskip~\cite{364506:8092431,364506:8092473}, Q-Chem\unskip~\cite{364506:8298308}, GAUSSIAN\unskip~\cite{364506:8298351,364506:8298350,364506:9918799}, MOLGW \unskip~\cite{364506:8298435,364506:8217217}, Quantum Espresso\unskip~\cite{364506:8298352}, and QBOX/QB@LL \unskip~\cite{364506:8093956,364506:9893475,364506:9918701}.  Amongst these codes, the implementations vary in the underlying basis sets used, with implementations ranging from real-space grids\unskip~\cite{364506:8298262,364506:8298307,364506:8092431}, Gaussian-type atomic orbitals \unskip~\cite{364506:8217306,364506:8298351,364506:8298350,364506:8298262,364506:8298435,364506:8217217}, plane-waves \unskip~\cite{364506:8093956}, and mixed Gaussians/planewaves\unskip~\cite{364506:8298304}.  

Recently, several RT-TDDFT implementations have exploited the gauge freedom inherent in the time-dependent Kohn Sham (TDKS) equations that underly RT-TDDFT calculations. Under arbitrary, unitary, gauge transformations, the physical properties of the electron dynamics are unchanged, but certain gauge transformations allow for reformulations of the TDKS equations that can be advantageous in certain physical and numerical situations \unskip~\cite{364506:8661882}. For instance, in recent works, Pemmaraju, et al. \unskip~\cite{364506:8092425} and Noda, et al. \unskip~\cite{364506:8298307} have implemented velocity-gauge formulations of RT-TDDFT that provide a convenient framework to simulation responses of periodic systems to both weak and intense electric fields which, in the velocity gauge, can be represented via vector potentials \unskip~\cite{364506:8661882}.  

Another interesting example of gauge transformations in RT-TDDFT, demonstrated in recent works by Lin and co-workers, is the propagation of electron dynamics in the parallel transport gauge \unskip~\cite{364506:8321071,364506:8321155}.  In the parallel transport gauge, the unitary transformation is performed such that oscillations of the propagating orbitals are minimized. This allows for significantly larger propagation time steps to be used with implicit time integration schemes such as the Crank-Nicolson scheme, providing computational advantages in practical RT-TDDFT simulations, especially when hybrid exchange-correlation functionals are involved \unskip~\cite{364506:8321072}.

In this work, we exploit gauge freedom in two regards: We represent the KS orbitals in the gauge of maximally localized Wannier functions (MLWFs), and, as it naturally follows, we represent electric fields in the length gauge corresponding to the formulation of finite fields as scalar potentials.  The foundational implementation of RT-TDDFT in the QB@LL branch of the Qbox code \unskip~\cite{364506:8093956,364506:9893475} used in this study involves a planewave pseudopotential formalism with periodic boundary conditions in which the TDKS states are represented as Bloch orbitals. However, in this work, through "on-the-fly" computation and application of a unitary transformation matrix at each RT-TDDFT simulation step, we propagate the TDKS orbitals as time-dependent MLWFs (TD-MLWFs).  Due to their relationship with the dynamic Berry phase \unskip~\cite{364506:8092420}, the TD-MLWFs allow us to calculate the dynamic polarization throughout RT-TDDFT simulations involving isolated and periodic systems in static or time-dependent electric fields.  Through such simulations, we can calculate linear response quantities such as optical absorption spectra, and we can also simulate nonlinear responses to strong fields associated with laser pulses or ionizing particle radiation.  Additionally, MLWFs are often used to give chemically intuitive representations of the electronic structure of molecules and materials in terms of bonds, lone-pairs, etc. and TD-MLWFs have the same utility in the context of RT-TDDFT.  By decomposing the dynamic polarization response in terms of contributions from TD-MLWFs associated with different chemical moieties, we determine, for instance, how a water molecule's lone-pair electrons contribute to individual peaks in the total optical absorption spectrum.  We also use the real-time propagation of MLWFs to study the electron excitation dynamics of a system in response to proton irradiation. Normally, excitations induced by photonic and ionic irradiation are difficult to compare, but our implementation of TD-MLWFs allows us to calculate a type of electronic stopping response spectrum, which can be used to make such comparisons. Finally, we perform simulations of optically-driven quantized charge transport in polyacetylene to demonstrate the length-gauge time-dependent field representation and to consider the possibilities for using TD-MLWF propagation to study dynamic topological phenomena.  
\bgroup
\fixFloatSize{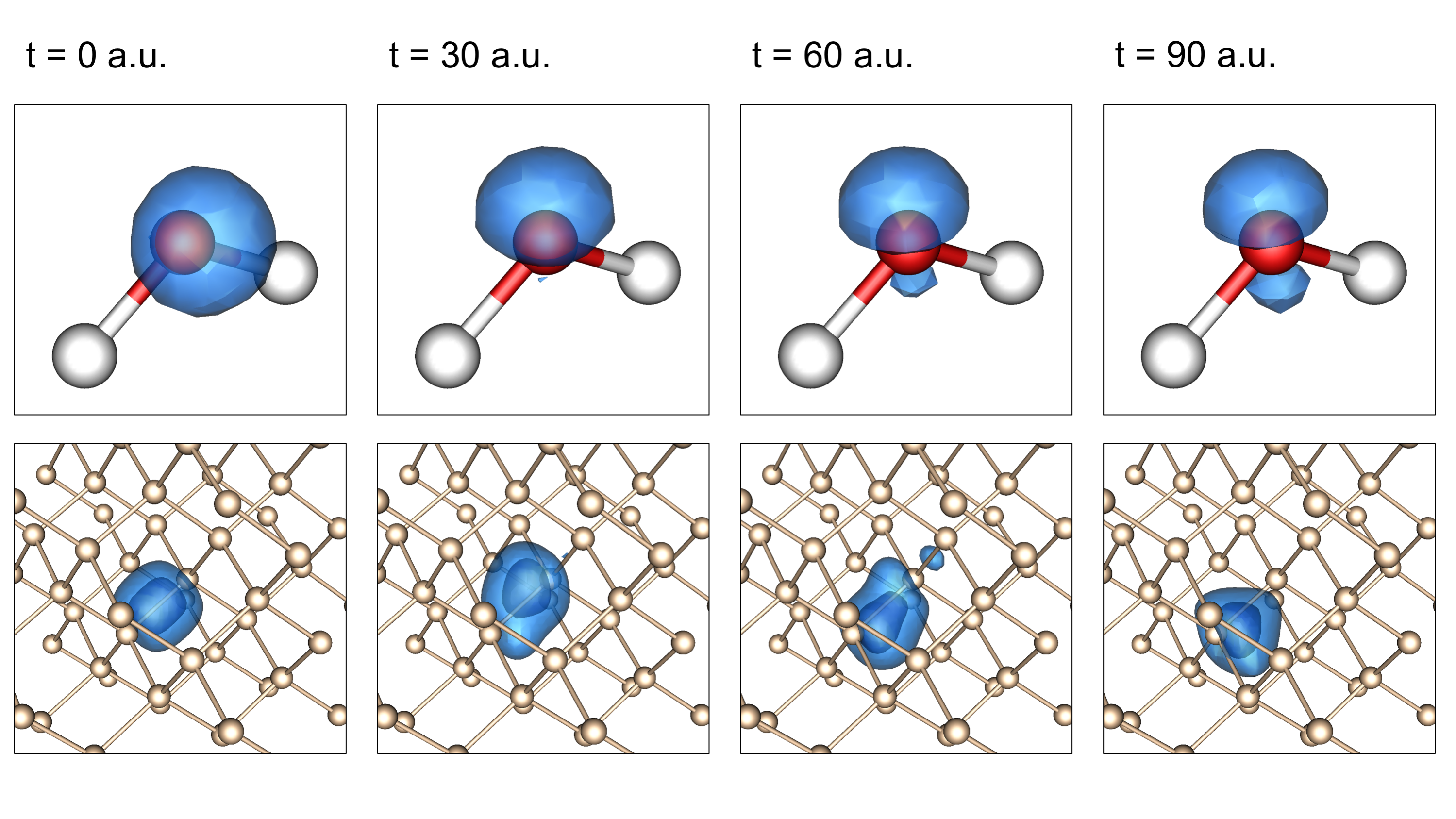}
\begin{figure*}[!htbp]
\centering \makeatletter\IfFileExists{images/e1ef6088-9760-49d0-a500-5eefe23ef223-ufig1.png}{\includegraphics[width=.74\linewidth]{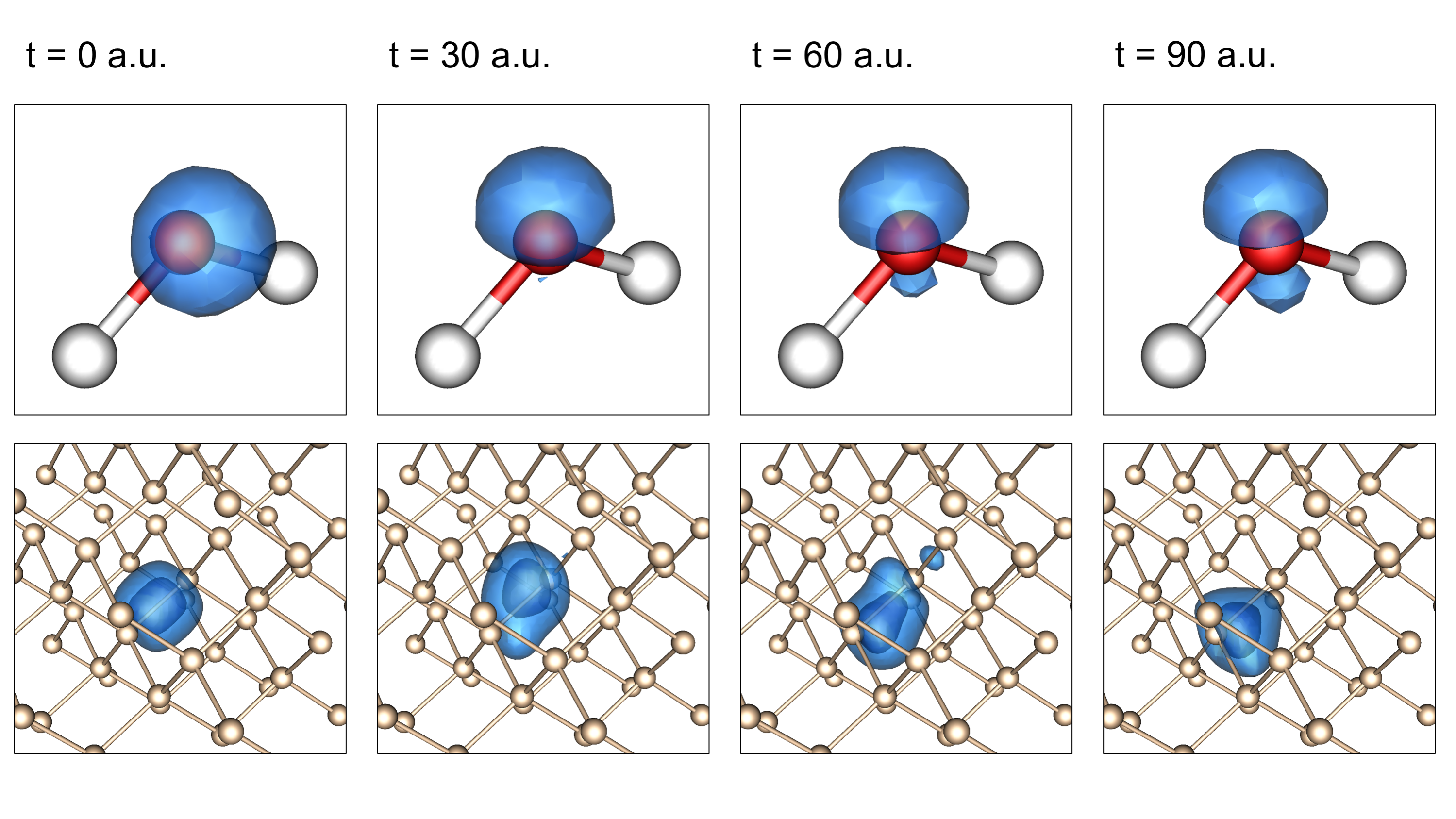}}{}
\makeatother 
\caption{{Isosurfaces of a single TD-MLWF orbital in an isolated water molecule (upper) and in crystalline silicon (lower). Each panel shows the TD-MLWFs at a different point in time in RT-TDDFT simulations after the systems have been perturbed by an instantaneous electric field impulse. }}
\label{f-b7fd}
\end{figure*}
\egroup

\section{Theoretical and Computational Details}

\subsection{Time-Dependent Kohn Sham Equations}The Runge-Gross theorem \unskip~\cite{364506:8296666} allows us to describe the quantum dynamics of a system evolving from a given initial state through a time-evolving electron probability density. This extension of the Hohenberg-Kohn theorem \unskip~\cite{364506:8093909}, formulated in the time domain, gave rise to the widely successful time-dependent density functional theory (TDDFT), now frequently used for excited state calculations and simulations. As in ground state DFT, practical TDDFT calculations rely on equations expressed in terms of a system of non-interacting Kohn-Sham (KS) particles in an effective KS potential.  The time-dependent Kohn-Sham equations (TDKS) are as follows:
\let\saveeqnno\theequation
\let\savefrac\frac
\def\dispfrac{\displaystyle\savefrac}
\begin{eqnarray}
\let\frac\dispfrac
\gdef\theequation{1}
\let\theHequation\theequation
\label{disp-formula-group-4702978655aa77e78f591a7b75d79771}
\begin{array}{@{}l}\begin{array}{ccl}i\hslash\frac d{dt}\vert\phi_i(\mathbf r,t)\rangle&=&\widehat H(t)\;\vert\phi_i(\mathbf r,t)\rangle\\&=&\left\{\widehat T+{\widehat V}_{ext}(t)+{\widehat V}_{HXC}\lbrack\rho\rbrack\right\}|\phi_i(\mathbf r,t)\rangle\end{array}\end{array}
\end{eqnarray}
\global\let\theequation\saveeqnno
\addtocounter{equation}{-1}\ignorespaces 

\let\saveeqnno\theequation
\let\savefrac\frac
\def\dispfrac{\displaystyle\savefrac}
\begin{eqnarray}
\let\frac\dispfrac
\gdef\theequation{2}
\let\theHequation\theequation
\label{disp-formula-group-81537fefcadc1bc54d4f4d63f5057b76}
\begin{array}{@{}l}\rho(\mathbf r\mathit,\mathit\;t)=\sum_i\left|\phi_i(\mathbf r\mathit,\mathit\;t)\right|^{2}\end{array}
\end{eqnarray}
\global\let\theequation\saveeqnno
\addtocounter{equation}{-1}\ignorespaces 
In Equation~(\ref{disp-formula-group-4702978655aa77e78f591a7b75d79771}), $\widehat T $ is the kinetic energy operator $\frac{-\hslash}{2m_{}}\nabla_\mathbf r^{2} $ and $V_{HXC}\lbrack\rho\rbrack(\mathbf r,t)=\int\frac{\rho(\mathbf r\boldsymbol',t)}{\left|\mathbf r-\mathbf r\boldsymbol'\right|}\operatorname d\mathbf r'+\frac{\delta E_{XC}}{\delta\rho(\mathbf r,t)} $  is the sum of the Hartree (H) potential and the exchange-correlation (XC) potential, which is derived from a universal XC functional $E_{XC}\lbrack\rho\rbrack $. The electron density $\rho $ here is expressed as the sum of square amplitudes of the occupied KS orbitals $\phi_i $ (labeled by the state index i). The total energy at time t can be expressed as
\let\saveeqnno\theequation
\let\savefrac\frac
\def\dispfrac{\displaystyle\savefrac}
\begin{eqnarray}
\let\frac\dispfrac
\gdef\theequation{3}
\let\theHequation\theequation
\label{disp-formula-group-ded4cf96e51008fd02badfc666c5da4b}
\begin{array}{@{}l}\begin{array}{l}\begin{array}{l}E(t)=\sum_i\left\langle\phi_i(\mathbf r,t)\left|\widehat T\right|\phi_i(\mathbf r,t)\right\rangle\\\;\;\end{array}\\+\int\rho(\mathbf r,t)V_{ext}(\mathbf r,t)\operatorname d\mathbf r\;+\;E_{HXC}\lbrack\rho\rbrack(t)\end{array}\end{array}
\end{eqnarray}
\global\let\theequation\saveeqnno
\addtocounter{equation}{-1}\ignorespaces 
This energy functional of the time-dependent density can be shown to obey  energy conservation when the so-called adiabatic approximation\unskip~\cite{364506:8093954} is adopted, which provides a useful observable for validating numerical implementations and practical simulations\unskip~\cite{364506:8093956}. In most RT-TDDFT simulations of realistic systems, the exchange-correlation (XC) potential is approximated with the adiabatic approximation. That is, the XC potential depends only on the instantaneous electron density, neglecting any memory effects.  While time-dependent XC functionals beyond the adiabatic approximation are an area of active research\unskip~\cite{364506:8094052,364506:8303647,364506:8303648,364506:8303649,364506:8303650}, such a topic is beyond the scope of this work. Even so, within the adiabatic approximation to the XC functional, TDDFT calculations has still been successful in predicting excited state properties for a diversity of molecular and material systems.

While the majority of applications of TDDFT remain within the linear-response formulation of TDDFT (LR-TDDFT), this real-time propagation approach (RT-TDDFT) has become increasingly popular in recent years.  In a RT-TDDFT calculation, the work entails solving a set of time-dependent KS equations, which are non-linear partial differential equations, in time, for given initial conditions. The time dependence of the external potential in the TDKS system can arise explicitly and implicitly from a variety of sources, depending on the phenomenon of interest.  For example, in applications of RT-TDDFT to study electronic stopping processes\unskip~\cite{364506:8217218,364506:8217261,364506:8217172,364506:8217219,364506:8297331,364506:9683361}\unskip~\cite{364506:9683403}, a fast-moving charged ion gives rise to an external potential that varies as the ion's position changes over the course of time.  In addition to ionic motion, the external potential can vary in time in situations involving a dynamic applied electric field.  A key advantage of RT-TDDFT is that it provides a framework within which a wide variety of dynamical electronic processes, such as electronic stopping and photoexcitation, can be treated on an equal footing.

\subsection{Bloch and Wannier Orbital Representations}In the KS and TDKS equations, a Bloch representation form naturally arises in extended systems, in which periodic boundary conditions are adapted\unskip~\cite{364506:8094142}.  The TDKS orbitals are written in the Bloch function form,
\let\saveeqnno\theequation
\let\savefrac\frac
\def\dispfrac{\displaystyle\savefrac}
\begin{eqnarray}
\let\frac\dispfrac
\gdef\theequation{4}
\let\theHequation\theequation
\label{disp-formula-group-68fdfe7d25a4e7123507dbba2e886786}
\begin{array}{@{}l}\varphi_{n\boldsymbol k}(\mathbf r,t)\;=u_{n\mathbf k}(\mathbf r,t)\;\;e^{i\mathbf k\boldsymbol\cdot\mathbf r}\end{array}
\end{eqnarray}
\global\let\theequation\saveeqnno
\addtocounter{equation}{-1}\ignorespaces 
where $u_{n\mathbf k}(\mathbf r,t) $ is the periodic part of the TDKS function, $n $ is the eigenstate index, and $\mathbf k $ is the Brillouin zone vector. By preserving the electron probability density from the KS single-particle orbitals, any unitary transformation of the above Bloch functions gives an equivalent physical description of the system.  Thus, one can apply a unitary transformation to the Bloch functions that results in a representation in a set of ``Wannier functions'' (WFs) \unskip~\cite{364506:8094143,364506:8094146,364506:8094189} as follows:
\let\saveeqnno\theequation
\let\savefrac\frac
\def\dispfrac{\displaystyle\savefrac}
\begin{eqnarray}
\let\frac\dispfrac
\gdef\theequation{5}
\let\theHequation\theequation
\label{disp-formula-group-3e524f1cf41e0efdbd3d8683a70ed03b}
\begin{array}{@{}l}w_{n\mathbf R}(\mathbf r,\mathrm t)\;=\;\frac\Omega{(2\pi)^{3}}\int_{BZ}d\mathbf k\boldsymbol\;e^{-i\mathbf k\cdot\mathbf R}\;\varphi_{\mathrm n\mathbf k}(\mathbf r,t)\end{array}
\end{eqnarray}
\global\let\theequation\saveeqnno
\addtocounter{equation}{-1}\ignorespaces 
where the Wannier functions $w_{n\mathbf R} $ are indexed with a band-like index $n $ located in the lattice-periodic cell $\mathbf R $ with a real-space cell volume $\Omega $.  There exists a gauge freedom in this procedure due to the fact that one can apply an arbitrary phase transformation $e^{i\theta_n(\mathbf k\boldsymbol)} $ to the Bloch orbitals without changing the physical observables of the system.  However, in terms of the resultant WFs, such a transformation to the Bloch orbitals could alter the shapes and spreads of the corresponding WFs.  Thus, due to this gauge freedom, in this most general definition, the WFs are not unique \unskip~\cite{364506:8092326}.

In order to address this nonuniqueness issue, a variety of schemes have been proposed, including projection methods \unskip~\cite{364506:8303815,364506:8303816,364506:8303817}, variational procedures \unskip~\cite{364506:8303863}, and methods based on symmetry considerations \unskip~\cite{364506:8303860,364506:8303861}. However, the most widely used method is the maximally localized Wannier function (MLWF) procedure developed by Marzari and Vanderbilt\unskip~\cite{364506:8303864}.  In this method, a spatial spread functional is defined, and then the unitary transformation can be optimized in order to minimize the spread functional, thus maximizing the localization of the WFs. Such maximally localized Wannier functions (MLWFs) typically have the character of being localized on chemical bonds and other familiar chemical moieties, which is why they have gained popularity for their use in interpreting the electronic structures of various matter \unskip~\cite{364506:8092326,364506:8297331}. Throughout the entirety of this work, MLWFs are used. The algorithmic and computational details of the localization procedure are discussed later in the manuscript.

Because the transformation between Bloch and MLWF states is unitary, both representations comprise equally valid descriptions of the electronic band subspace \unskip~\cite{364506:8092326}. Thus, in the TDKS equations, one can substitute the single-particle KS orbitals, normally taken to be the Bloch functions, with MLWFs.  The charge density obtained by summing the square amplitudes of the KS orbitals is identical to that obtained by summing the square amplitudes of the MLWFs, meaning that we have an equivalent framing of RT-TDDFT calculations in terms of time-dependent MLWFs (TD-MLWFs). The TDKS equation in Bloch representation is 
\let\saveeqnno\theequation
\let\savefrac\frac
\def\dispfrac{\displaystyle\savefrac}
\begin{eqnarray}
\let\frac\dispfrac
\gdef\theequation{6}
\let\theHequation\theequation
\label{dfg-f54c}
\begin{array}{@{}l}\style{font-size:12px}{\begin{array}{ccl}i\frac d{dt}\vert u_{j,\mathbf k}(\mathbf r,t)\rangle&=&\left\{\frac1{2m}\left[-i\hslash{\boldsymbol\nabla}_\mathbf r+\mathbf k\right]^{2}\;\right.+\;{\widehat V}_{ion}(t)\\&&\left.+\;{\widehat V}_{HXC}\lbrack\{\;{u_{j,\mathbf k}(\mathbf r,t)}\}\rbrack\right\}\vert u_{j,\mathbf k}(\mathbf r,t)\rangle\end{array}}\end{array}
\end{eqnarray}
\global\let\theequation\saveeqnno
\addtocounter{equation}{-1}\ignorespaces 
where this equation can also be written in terms of Wannier functions as
\let\saveeqnno\theequation
\let\savefrac\frac
\def\dispfrac{\displaystyle\savefrac}
\begin{eqnarray}
\let\frac\dispfrac
\gdef\theequation{7}
\let\theHequation\theequation
\label{dfg-4e98db1f354d}
\begin{array}{@{}l}\style{font-size:12px}{\begin{array}{ccl}i\frac d{dt}\vert w_l(\mathbf r,t)\rangle&=&\left\{\widehat U^{ML}+\frac1{2m}\left[-i\hslash{\boldsymbol\nabla}_\mathbf r\right]^{2}\;\right.+\;{\widehat V}_{ion}(t)\\&&\left.+\;{\widehat V}_{HXC}\lbrack\{\;{w_l(\mathbf r,t)}\}\rbrack\right\}\vert w_l(\mathbf r,t)\rangle\end{array}}\end{array}
\end{eqnarray}
\global\let\theequation\saveeqnno
\addtocounter{equation}{-1}\ignorespaces 
where ${\style{font-size:12px}w}_\style{font-size:12px}l\style{font-size:12px}(\style{font-size:12px}{\mathbf r}\style{font-size:12px},\style{font-size:12px}t\style{font-size:12px}) $ is the time-dependent MLWF (TD-MLWF) with a band-like index $l $, where $\widehat U^{ML} $ is the unitary operator that ensures the maximal localization:
\let\saveeqnno\theequation
\let\savefrac\frac
\def\dispfrac{\displaystyle\savefrac}
\begin{eqnarray}
\let\frac\dispfrac
\gdef\theequation{8}
\let\theHequation\theequation
\label{dfg-3585}
\begin{array}{@{}l}\widehat U^{ML}\vert w_l(t)\rangle\equiv\sum_m^{N}U_{lm}(t)\vert w_m(t)\rangle\end{array}
\end{eqnarray}
\global\let\theequation\saveeqnno
\addtocounter{equation}{-1}\ignorespaces 
where the unitary transformation matrix $U_{lm} $ minimizes the spread functional
\let\saveeqnno\theequation
\let\savefrac\frac
\def\dispfrac{\displaystyle\savefrac}
\begin{eqnarray}
\let\frac\dispfrac
\gdef\theequation{9}
\let\theHequation\theequation
\label{dfg-d464}
\begin{array}{@{}l}Min{\left\{\sum_n^{N}\left[\left\langle w_n\left|\widehat{\mathbf r}^{2}\right|w_n\right\rangle-\left\langle w_n\left|\widehat{\mathbf r}\right|w_n\right\rangle^{2}\right]\right\}}_U\end{array}
\end{eqnarray}
\global\let\theequation\saveeqnno
\addtocounter{equation}{-1}\ignorespaces 
where the quantum mechanical position operator $\widehat{\mathbf{r}} $ is defined generally even for extended systems\unskip~\cite{364506:8092421};
\let\saveeqnno\theequation
\let\savefrac\frac
\def\dispfrac{\displaystyle\savefrac}
\begin{eqnarray}
\let\frac\dispfrac
\gdef\theequation{10}
\let\theHequation\theequation
\label{dfg-8dc1}
\begin{array}{@{}l}\left\langle\widehat{\mathbf{r}}\right\rangle=\frac L{2\ensuremath{\pi}}Im\;\ln\left\langle\psi\left|e^{\frac{\mathrm i2\ensuremath{\pi}}{\mathrm L}\widehat{\mathbf{r}}}\right|\psi\right\rangle\end{array}
\end{eqnarray}
\global\let\theequation\saveeqnno
\addtocounter{equation}{-1}\ignorespaces 
where $L=\left\ensuremath{\Vert}\mathbf R\right\ensuremath{\Vert} $ is the cell dimension. The unitary operator inEquation~(\ref{dfg-4e98db1f354d}) does not change the quantum dynamics governed by the time-dependent electron density, and therefore it does not affect physical observables. The algorithmic details of the unitary transformation matrix calculation are discussed later in the "Implementation in QB@LL" section. 

One of the central motivations for carrying out this transformation to MLWFs is that it allows for the calculation of electric polarization in extended periodic systems.  While simple formulas exist for the calculation of electric dipoles of molecular systems with localized wavefunctions, these formulas cannot generally be extended to periodic systems in which the Bloch wavefunctions are delocalized in real space.  In extended systems, the analog to the electric dipole moment is the electric polarization, defined as the electric dipole moment per unit volume. While the calculation of this quantity for periodic systems may seem intuitive, the arbitrary choice between different valid unit cells can result in contradictory results for the electric polarization. This infamous problem was elegantly solved by Resta\unskip~\cite{364506:8303909} and King-Smith and Vanderbilt \unskip~\cite{364506:8303907} in work that is now collectively known as the ``modern theory of polarization'' \unskip~\cite{364506:8303910}.  In the modern theory of polarization, the electric polarization of a periodic system can be formulated in terms of Berry phases, or equivalently, in terms of Wannier functions. 

MLWFs have been used in the context of ground-state DFT to calculate the adiabatic stationary resonant state of an insulating material in static electric fields \unskip~\cite{364506:8092428,364506:8093508}. Souza, et al. also showed that this concept could be extended to dynamics via the definition of a nonadiabatic Berry phase polarization \unskip~\cite{364506:8092420}. Thus, through the incorporation of MLWFs in RT-TDDFT we can track the dynamics of a system's polarization in time.  Access to the dynamic polarization of a system in response to an external perturbation allows for the calculation of important quantities such as the transient current\textbf{\space }and frequency-dependent optical absorption \unskip~\cite{364506:8092420}.

\subsection{Finite Electric Fields and Periodic Boundary Conditions}While it is relatively straight-forward to perform RT-TDDFT calculations with time-varying external potentials caused by ionic motion, as has been done in RT-TDDFT studies of electronic stopping \unskip~\cite{364506:8217218,364506:8217172}, the treatment of spatially homogeneous electric fields requires more careful consideration in cases involving periodic systems. Fundamentally speaking, application of the Runge Gross theorem in cases of an extended periodic system in a homogeneous electric field is not formally justified. Instead, time-dependent \textit{current}-density functional theory (TDCDFT) should be used due to its incorporation of the macroscopic current\unskip~\cite{364506:8095027}. Despite this formal limitation,  RT-TDDFT studies have shown that many response properties can in practice be successfully acquired without TDCDFT in practice.  That being said, there are several additional theoretical complications to be considered. One way to treat response of electrons to electric field excitations in RT-TDDFT is to introduce time-dependent spatially-homogeneous electric field in the KS Hamiltonian through a scalar potential in the so-called "length gauge".  The electric field can be included through an additional potential in the Hamiltonian
\let\saveeqnno\theequation
\let\savefrac\frac
\def\dispfrac{\displaystyle\savefrac}
\begin{eqnarray}
\let\frac\dispfrac
\gdef\theequation{11}
\let\theHequation\theequation
\label{disp-formula-group-0698684ea59e828a45660038b12ddd36}
\begin{array}{@{}l}\style{font-size:12px}{{\widehat V}_\mathbf E(t)=\;e\mathbf E(t)\cdot\widehat{\mathbf r}\;}\end{array}
\end{eqnarray}
\global\let\theequation\saveeqnno
\addtocounter{equation}{-1}\ignorespaces 
where $\mathbf E(t) $and $\widehat{\mathbf{r}} $ are the time-dependent electric field and the quantum-mechanical position operator, respectively. However, the added potential makes the Hamiltonian incompatible with periodic boundary conditions (PBC) as needed for modeling extended systems. Thus, instead of this length gauge formulation, it is common to move to a different gauge in electromagnetism. The electric field can be equivalently represented as the magnetic flux by the vector potential.
\let\saveeqnno\theequation
\let\savefrac\frac
\def\dispfrac{\displaystyle\savefrac}
\begin{eqnarray}
\let\frac\dispfrac
\gdef\theequation{12}
\let\theHequation\theequation
\label{disp-formula-group-c0c084e97ea6e3ea3c01544e06f72318}
\begin{array}{@{}l}\mathbf A(t)=-c\int_{}^{t}\mathbf E(t')\operatorname dt'\end{array}
\end{eqnarray}
\global\let\theequation\saveeqnno
\addtocounter{equation}{-1}\ignorespaces 
In literature, RT-TDDFT simulations of periodic systems with homogeneous electric field employ this so-called velocity-gauge formulation \unskip~\cite{364506:8661882,364506:8842990,364506:8092425,364506:8351790}.  The TDKS equation, can be written as 
\let\saveeqnno\theequation
\let\savefrac\frac
\def\dispfrac{\displaystyle\savefrac}
\begin{eqnarray}
\let\frac\dispfrac
\gdef\theequation{13}
\let\theHequation\theequation
\label{disp-formula-group-8d061b1910fa64e848a987eb3994f94e}
\begin{array}{@{}l}\style{font-size:14px}{\begin{array}{ccc}i\hslash\frac d{dt}\vert u_{i,\mathbf k}(\mathbf r,t)\rangle&=&\left\{\frac1{2m}\left[-i\hslash{\boldsymbol\nabla}_\mathbf r+\mathbf k\boldsymbol+\frac ec\mathbf A(t)\right]^{2}\right.\\&&\left.+{\widehat V}_{ext}(t)+{\widehat V}_{HXC}\lbrack\rho\rbrack\right\}\vert u_{i,\mathbf k}(\mathbf r,t)\rangle\end{array}}\end{array}
\end{eqnarray}
\global\let\theequation\saveeqnno
\addtocounter{equation}{-1}\ignorespaces 
In the above velocity gauge, the Hamiltonian associated with Equation~(\ref{disp-formula-group-8d061b1910fa64e848a987eb3994f94e}) is periodic for spatially homogeneous electric fields, allowing for a Bloch wavefunction to be used in the TDKS equations \unskip~\cite{364506:8092425}. Although the velocity gauge is commonly used when simulating extended periodic systems, it is not without its limitations. In particular, unlike the scalar potential, the vector-potential results in a Hamiltonian which inherently changes nonadiabatically in time, even for the static field case\unskip~\cite{364506:8092420}.  Thus, unlike the length gauge, the velocity-gauge is not suitable for calculating the stationary resonant polarization state of an insulator in a static electric field using time-independent DFT. 

The use of MLWFs instead of Bloch orbitals allow us to employ the length gauge in which the homogeneous electric field is represented by a scalar potential in the Hamiltonian.  For non-metallic systems with a finite band gap (termed "Wannier-representable" systems \unskip~\cite{364506:8092420}), the MLWFs are spatially well-localized within each periodic simulation cell, and the scalar potential can be applied to each MLWF individually such that the same homogeneous electric field is described. This formalism been already demonstrated for the static case\unskip~\cite{364506:9153313} in the context of first-principles molecular dynamics simulations \unskip~\cite{364506:9422764}. \textbf{\space }Due to their connection to the dynamic Berry phase polarization (discussed in more detail in the next section), MLWFs provide a computationally attractive, and physically intuitive, avenue for simulating systems in finite electric fields with RT-TDDFT in the MLWF/length-gauge. This TDKS/TD-MLWF equation can be written as
\let\saveeqnno\theequation
\let\savefrac\frac
\def\dispfrac{\displaystyle\savefrac}
\begin{eqnarray}
\let\frac\dispfrac
\gdef\theequation{14}
\let\theHequation\theequation
\label{dfg-a8c4}
\begin{array}{@{}l}\style{font-size:11px}{\begin{array}{ccl}i\hslash\frac d{dt}\vert w_i(\mathbf r,t)\rangle&=&\begin{array}{cc}\left\{\widehat U^{ML}+\left[\frac{\style{font-size:12px}1}{\style{font-size:12px}2\style{font-size:12px}m}\left[\style{font-size:12px}-\style{font-size:12px}i\style{font-size:12px}\hbar{\style{font-size:12px}{\boldsymbol\nabla}}_\style{font-size:12px}{\mathbf r}\right]^\style{font-size:12px}2+{\widehat V}_{ext}(t)\right.\right.&\end{array}\\&&\left.\left.+{\widehat V}_{HXC}\lbrack\rho\rbrack+e\mathbf E(t)\cdot\widehat r\;\right]\right\}\vert w_i(\mathbf r,t)\rangle\end{array}}\end{array}
\end{eqnarray}
\global\let\theequation\saveeqnno
\addtocounter{equation}{-1}\ignorespaces 
Including the unitary operator $\widehat U^{ML} $ here, as in Equation~(\ref{dfg-4e98db1f354d}), ensures that the Wannier functions remain maximally localized during the propagation, and this allows us to use the length gauge for applying a homogeneous electric field even when periodic boundary conditions are adapted.  In practice, preserving the maximal locality of the Wannier functions enable us to easily calculate the positions of the MLWF centers at each time step via the diagonal elements of the quantum-mechanical position operator matrices (discussed in more detail in the "Maximal Localization Procedure". See Equation~(\ref{disp-formula-group-49dc00b5a337833868e25f912d87190a})).

\subsection{Dynamic Polarization via Wannier Functions}The properties of insulating crystals in static electric fields can be calculated via the iterative determination of field-dependent WFs through the minimization of a so-called "electric enthalpy" functional \unskip~\cite{364506:8092428}. This formalism was also generalized and extended to the time-dependent domain in work by Souza, et al. \unskip~\cite{364506:8092420} in which it was shown that the dynamic polarization can be expressed as a nonadiabatic geometric phase. The dynamic current is defined as the rate of polarization change per unit volume with respect to time:
\let\saveeqnno\theequation
\let\savefrac\frac
\def\dispfrac{\displaystyle\savefrac}
\begin{eqnarray}
\let\frac\dispfrac
\gdef\theequation{15}
\let\theHequation\theequation
\label{disp-formula-group-984e5b25039f9f5df0be1f6afa7b85ac}
\begin{array}{@{}l}\mathbf J\boldsymbol\;(t)=\frac{d{\mathbf P}_{el}(t)}{dt}\end{array}
\end{eqnarray}
\global\let\theequation\saveeqnno
\addtocounter{equation}{-1}\ignorespaces 
Where the dynamic electronic polarization  ${\mathbf P}_{el}(t) $ is given by the valence-band Berry phase \unskip~\cite{364506:8093457,364506:8092428,364506:8303907}
\let\saveeqnno\theequation
\let\savefrac\frac
\def\dispfrac{\displaystyle\savefrac}
\begin{eqnarray}
\let\frac\dispfrac
\gdef\theequation{16}
\let\theHequation\theequation
\label{disp-formula-group-7ea75e015136d52d281fd09fb69b64e5}
\begin{array}{@{}l}\style{font-size:10px}{\begin{array}{l}\begin{array}{l}{\mathbf P}_{el}(t)=\\\\\frac{-2e}{(2\pi)^{3}}\sum_n\int_{BZ}^{}\operatorname d\mathbf k\;\left\langle u_{n,\mathbf k}(\mathbf r,t)\left|i\nabla_\mathbf k\right|u_{n,\mathbf k}(\mathbf r,t)\right\rangle\end{array}\\\end{array}}\end{array}
\end{eqnarray}
\global\let\theequation\saveeqnno
\addtocounter{equation}{-1}\ignorespaces 
Where $u_{n,\mathbf k} $ are the Bloch functions characterized by a band index $n $. In this time-dependent case, we do not assume that changes in the Hamiltonian are adiabatic. Instead, Equation~(\ref{disp-formula-group-7ea75e015136d52d281fd09fb69b64e5}) can be interpreted as the nonadiabatic geometric phase \unskip~\cite{364506:8092420}.  Alternatively, this Berry phase expression can be transformed into a real-space representation in terms of the occupied Wannier functions such as MLWFs:
\let\saveeqnno\theequation
\let\savefrac\frac
\def\dispfrac{\displaystyle\savefrac}
\begin{eqnarray}
\let\frac\dispfrac
\gdef\theequation{17}
\let\theHequation\theequation
\label{dfg-ec9f}
\begin{array}{@{}l}\style{font-size:11px}{{\mathbf P}_\mathbf{el}(t)\;=\;-\frac{2e}\Omega\sum_i\left\langle w_i(\mathbf r\boldsymbol-\mathbf R,t)\left|\widehat{\mathbf r}\right|w_i(\mathbf r\boldsymbol-\mathbf R,t)\right\rangle}\end{array}
\end{eqnarray}
\global\let\theequation\saveeqnno
\addtocounter{equation}{-1}\ignorespaces 
where the dynamic polarization is recast in real space as the vector sum of the charge centers of mass of the MLWFs (MLWFCs), $w_i $ which correspond to the expectation values of the quantum mechanical position operator $\widehat{ \mathbf{r}} $ for a periodic system. Here, the location of the MLWFC is indeterminate modulus the lattice vector $\mathbf R $, and consequently the dynamic Berry-phase polarization ${\mathbf P}_\mathrm{el}\boldsymbol(t) $ is also indeterminate mod $-\frac{2e\mathbf R}\Omega $.  This uncertainty is the so-called "quantum of polarization". Thus, as described in the modern theory of polarization, it is actually the change in polarization that is the physical quantity that needs to be measured.  

Using MLWFs, one can obtain this physically intuitive definition of the polarization, which is expressed in terms of the geometric centers of charge of the MLWFs (often called ``Wannier centers'', or WCs). As illustrated by Equation~(\ref{dfg-ec9f}), the dynamic current is then proportional to the displacement of the WCs.  Although the dynamic polarization is also gauge invariant only up to a "quantum of polarization", the current $\mathbf J(t) $ is uniquely defined. In fact, it is the current $\mathbf J(t) $ which is the quantity that can be used for determining various properties of the system of interest as discussed next.

The theoretical framework laid out above is general and allows for arbitrarily strong and rapid variations of the homogeneous electric field. Consequently, one can perform RT-TDDFT calculations to study systems (both molecular and extended) under photo irradiation (i.e., photo excitation), ion irradiation (i.e., electronic stopping), and static electric fields on an equal footing.  As mentioned earlier, it should be noted that the underlying derivation of the dynamic polarization as a nonadiabatic Berry phase requires that the initial state be "Wannier-representable" (WR), as described in Ref. \unskip~\cite{364506:8092420}. Physically, Wannier-representability holds when the initial state of the system is insulating-like, not metallic, in the RT-TDDFT simulations. Additionally, In their work, Souza, et al. proved that, in the absence of scattering, a WR state remains WR or "insulating like" at all later times, even if the ground state of the Hamiltonian for a given ionic configuration becomes metallic.

With access to dynamic current, one can obtain the frequency-dependent conductivity and also dielectric function within linear response theory. For a system under a homogeneous perturbing field $E(t) $ in the $\nu $-direction, and for the time-dependent current in the $\mu $-direction, the frequency dependent conductivity is obtained as 
\let\saveeqnno\theequation
\let\savefrac\frac
\def\dispfrac{\displaystyle\savefrac}
\begin{eqnarray}
\let\frac\dispfrac
\gdef\theequation{18}
\let\theHequation\theequation
\label{disp-formula-group-05a7ec00aab60effb08e925c0363f14b}
\begin{array}{@{}l}\sigma _{\mu\nu}\left(\omega\right)\;=\;\frac1{\widetilde{E_\nu}(\omega)}\int_{}^{T}dt\;e^{i\omega t}\;J_\mu(t)\end{array}
\end{eqnarray}
\global\let\theequation\saveeqnno
\addtocounter{equation}{-1}\ignorespaces 
where $\widetilde E(\omega) $ is the Fourier transform of the applied electric field.  For extended periodic systems, the frequency dependent dielectric function is related to the conductivity via
\let\saveeqnno\theequation
\let\savefrac\frac
\def\dispfrac{\displaystyle\savefrac}
\begin{eqnarray}
\let\frac\dispfrac
\gdef\theequation{19}
\let\theHequation\theequation
\label{disp-formula-group-c9e06f68e91e083233a03d0eb76786b5}
\begin{array}{@{}l}\varepsilon(\omega)\;=\;1\;+\;\frac{4\pi\;i\;}{3\omega}Tr\;\lbrack\sigma _{\mu\nu}\left(\omega\right)\rbrack\end{array}
\end{eqnarray}
\global\let\theequation\saveeqnno
\addtocounter{equation}{-1}\ignorespaces 
where $Tr\lbrack\sigma (\omega)\rbrack $ is the trace of the complex conductivity tensor.  For extended systems, the imaginary part of this dielectric function is directly related to the optical absorption, whereas the real part is related to dispersion.  For isolated systems, the macroscopic dielectric function is not well-defined, and instead the convention is to describe optical absorption in terms of the dipole strength function: 
\let\saveeqnno\theequation
\let\savefrac\frac
\def\dispfrac{\displaystyle\savefrac}
\begin{eqnarray}
\let\frac\dispfrac
\gdef\theequation{20}
\let\theHequation\theequation
\label{disp-formula-group-936a4dbe9b49a2a383f4d1f203d6b46d}
\begin{array}{@{}l}S(\omega)\;=\;\frac{4\pi\omega}{3c}Tr\;\lbrack\;Im\;\sigma _{\mu\nu}\left(\omega\right)\rbrack\end{array}
\end{eqnarray}
\global\let\theequation\saveeqnno
\addtocounter{equation}{-1}\ignorespaces 
where $\sigma _{\mu\nu}(\omega) $ is generally referred to as the frequency-dependent polarizability in the case of isolated systems \unskip~\cite{364506:9893431}. For computing absorption spectra using RT-TDDFT, one must choose an appropriate excitation procedure that simultaneously excites the system in a superposition of eigenstates. Any sudden perturbation at time t=0 that is suddenly "switched off" at the next time step has the effect of inducing electronic oscillations in the system that includes all frequency components and thus results in a broadband electronic excitation of the system \unskip~\cite{364506:8839262}. In RT-TDDFT simulations, there are two common choices for this sudden perturbation: First, there is the delta-function-like impulsive electric field.  In this approach, a finite electric field is applied only for an infinitesimally small moment in time. In practice, however, RT-TDDFT calculations involve a finite time-step $\Delta t $, which allows for an approximation of a true impulsive electric field. This has consequences for the electric field Fourier transform term  $\widetilde E(\omega) $, which for a true impulsive field with magnitude $E_0 $ should yield $\widetilde E(\omega)\;=\;E_0 $.  The equations defining the temporal profile and the Fourier transform of the electric field, with some magnitude $E_0 $, is defined below for the impulse approach: 
\let\saveeqnno\theequation
\let\savefrac\frac
\def\dispfrac{\displaystyle\savefrac}
\begin{eqnarray}
\let\frac\dispfrac
\gdef\theequation{21}
\let\theHequation\theequation
\label{dfg-88410796faf6}
\begin{array}{@{}l}\begin{array}{l}\begin{array}{l}E_{imp}(t)\;=\;\left\{\begin{array}{lc}E_0&\mathrm{for}\;t\;=\;0\\0&\mathrm{for}\;t\;>\;0\end{array}\right.\\{\widetilde E}_{imp}(\omega)=\;E_0\end{array}\end{array}\end{array}
\end{eqnarray}
\global\let\theequation\saveeqnno
\addtocounter{equation}{-1}\ignorespaces 
In practice, however, due to the finite integration time step, the field profile is actually that of a "boxcar" function, and its Fourier transform results in a sinc function.  An additional complication is that this approach can cause some difficulties in the numerical integration of the TDKS equations \unskip~\cite{364506:8839262}. For these reasons, in this work, we primarily use a second, more numerically convenient "step function" electric field approach, proposed by Yabana, et al. \unskip~\cite{364506:8093509}.  In this alternative method, one performs a standard DFT calculation including a static uniform electric field to acquire a stationary state solution which is subsequently propagated in the TDKS equations for t {\textgreater} 0 without the field.  This amounts to an adiabatic "switching on" of the field and a nonadiabatic "switching off" of the field at t=0 of the RT-TDDFT simulation.  The equations defining the temporal profile and the Fourier transform of the electric field, with some magnitude $E_0 $, is defined below for the step function approach: 
\let\saveeqnno\theequation
\let\savefrac\frac
\def\dispfrac{\displaystyle\savefrac}
\begin{eqnarray}
\let\frac\dispfrac
\gdef\theequation{22}
\let\theHequation\theequation
\label{dfg-5c282402a4b5}
\begin{array}{@{}l}\begin{array}{l}E_{step}(t)=\left\{\begin{array}{lc}E_0&\mathrm{for}\;t<0\\0&\mathrm{for}\;t>0\end{array}\right.\\{\widetilde E}_{step}(\omega)\;=\;\frac{E_0}{i\omega}\end{array}\end{array}
\end{eqnarray}
\global\let\theequation\saveeqnno
\addtocounter{equation}{-1}\ignorespaces 
In practice, of course, for a finite amount of time $T $. Consequently, the abrupt end of the oscillations in the polarization at the end of the numerical simulation leads to artificial ``wiggles'' in the calculated spectrum. The prominence of such numerical artifacts can be lessened to some degree by employing a damping function in the Fourier transform. A simple choice for the damping function, which we employ in this work, is the damping function $f(t)=e^{-\gamma t} $ where $\gamma $ is a damping constant.  For RT-TDDFT simulations converged sufficiently with respect to total simulation time, the application of this damping function results spectra with Lorentzian shaped peaks.

\subsection{Implementation in QB@LL}
\bgroup
\fixFloatSize{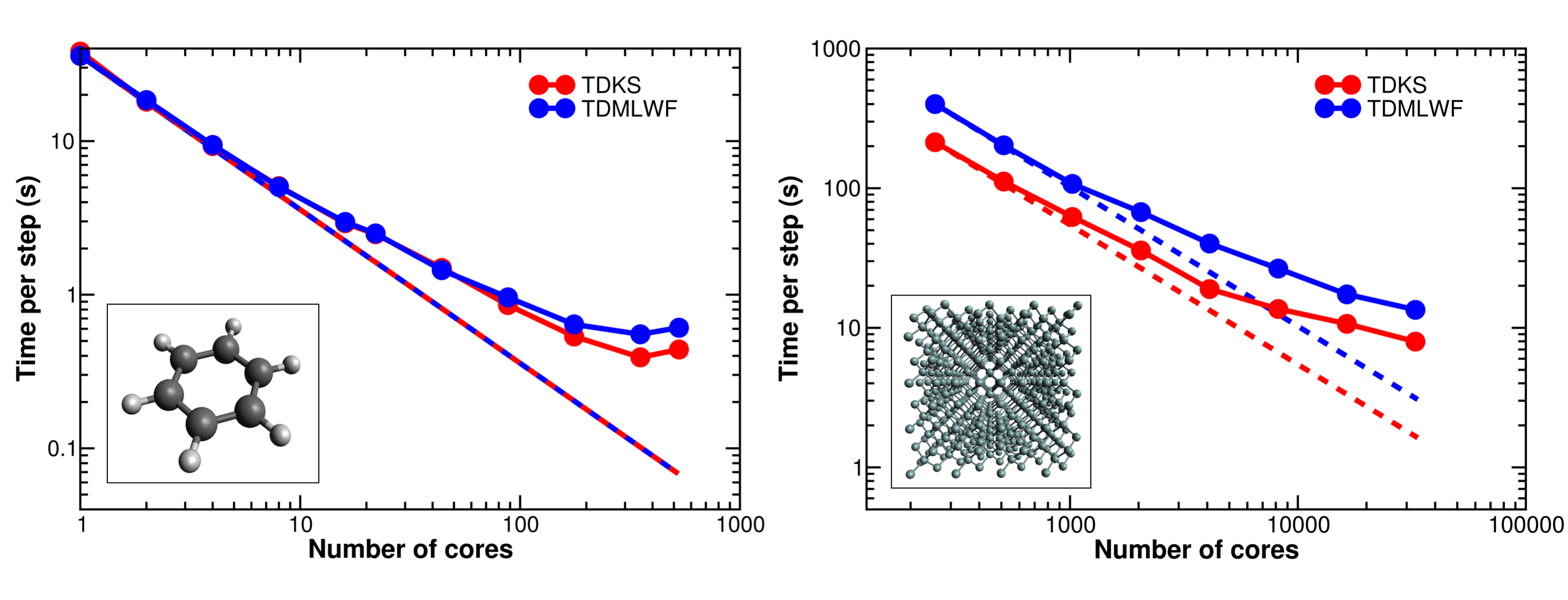}
\begin{figure*}[!htbp]
\centering \makeatletter\IfFileExists{images/b6fea859-84ae-4343-8d43-020f68036c55-ufig2.png}{\includegraphics{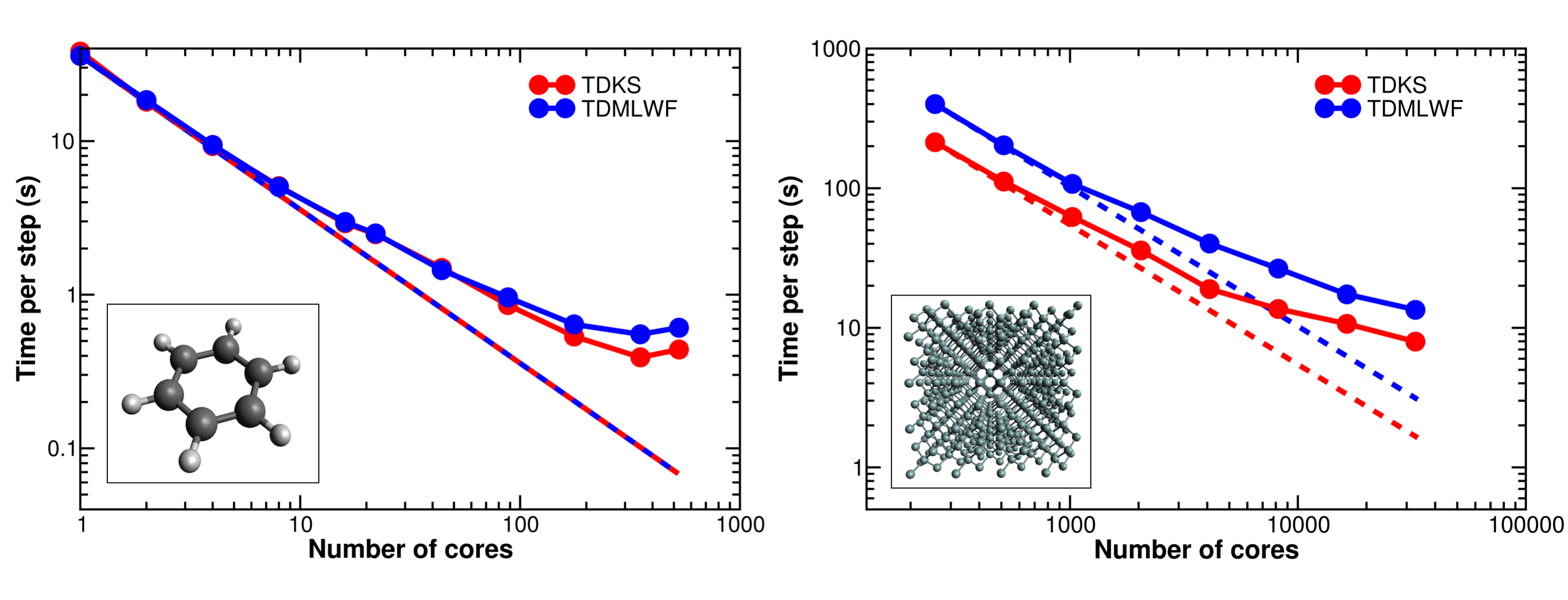}}{}
\makeatother 
\caption{{Performance results, indicated in time per RT-TDDFT step, showing the scalability of the TD-MLWF propagation (blue) versus the standard TDKS propagation (red) in RT-TDDFT simulations, for an isolated benzene molecule (left) and 512-atom crystalline silicon supercell (right). }}
\label{f-33edac16ff73}
\end{figure*}
\egroup

\bgroup
\fixFloatSize{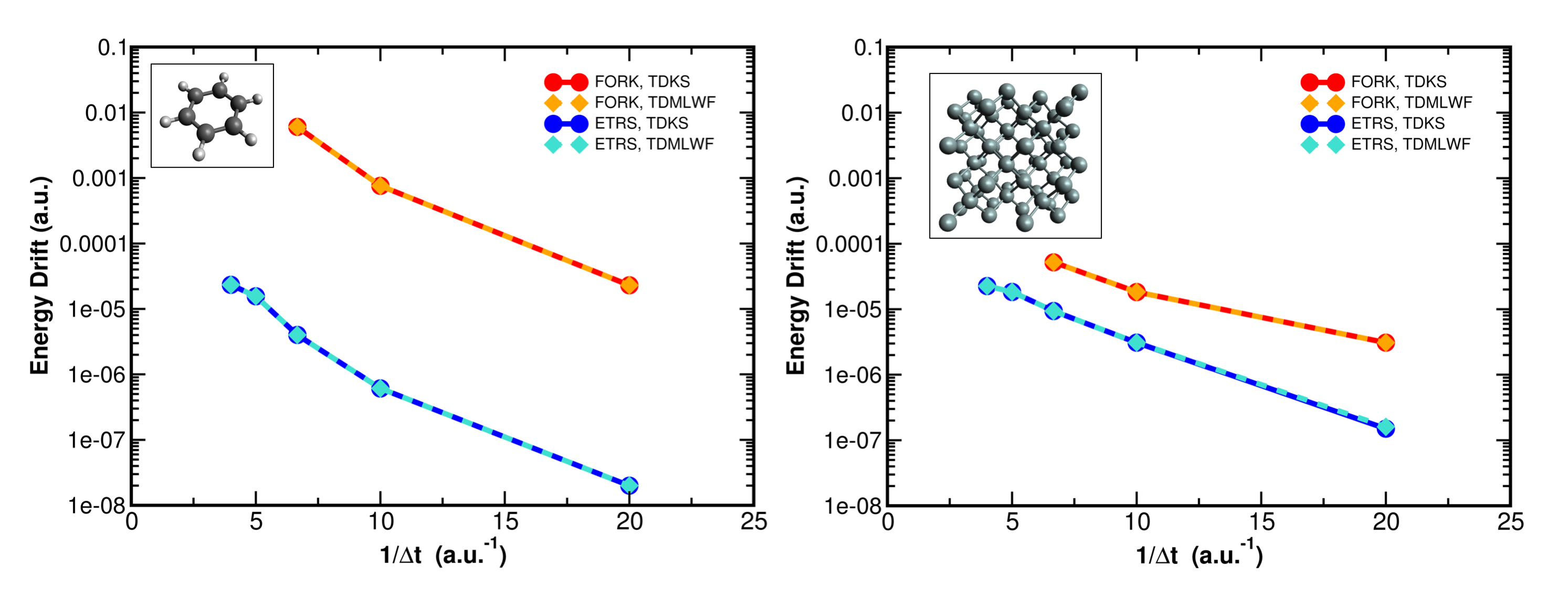}
\begin{figure*}[!htbp]
\centering \makeatletter\IfFileExists{images/3d13221b-89ab-4c18-a427-9f99b7a9520b-ufig3.png}{\includegraphics{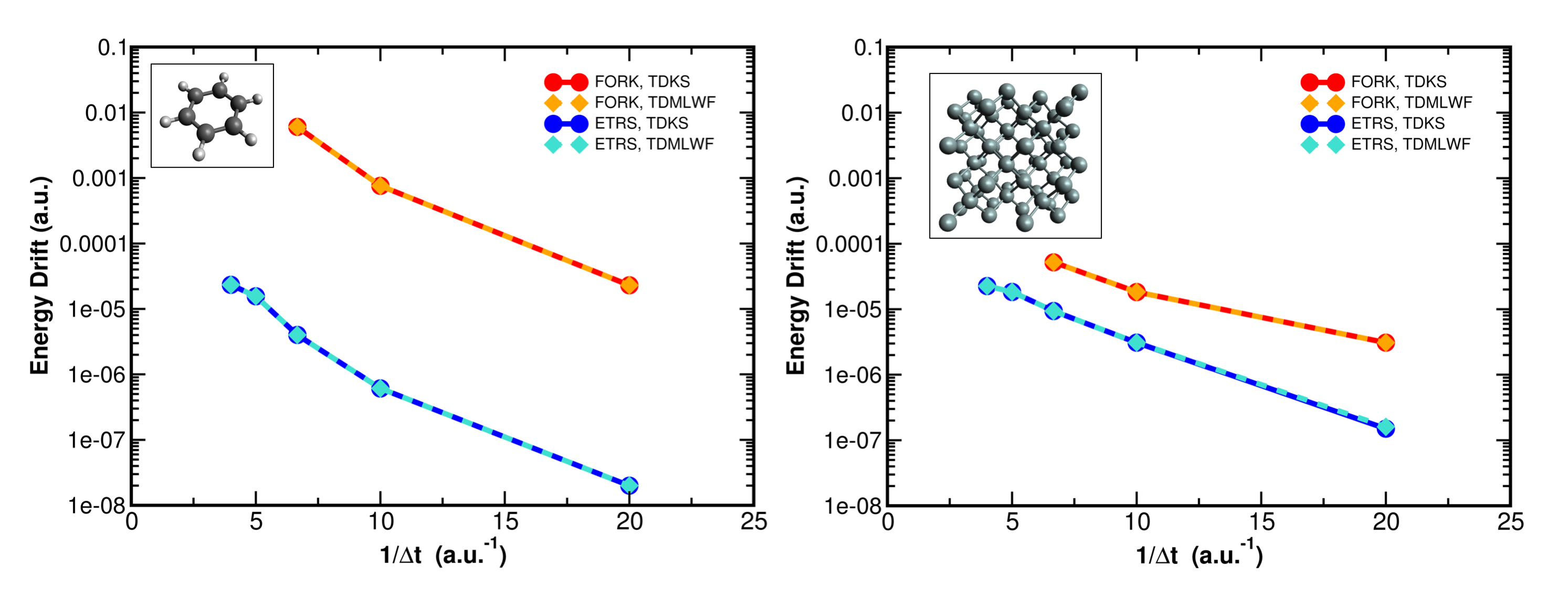}}{}
\makeatother 
\caption{{RT-TDDFT simulation accuracy, measured in terms of energy drift, for simulations with two different integration schemes (ETRS and FORK) and two different types of orbital propagation (TD-MLWF and TDKS), as a function of number of integrations per a.u. of time for an isolated benzene molecule (left), and 64-atom crystalline silicon supercell (right). }}
\label{f-c9a1669663a1}
\end{figure*}
\egroup
In this work, we employ the plane-wave pseudopotential formalism in RT-TDDFT as discussed in Ref. \unskip~\cite{364506:8093956}. The simulations carried out in this work were performed on the highly parallelized plane-wave pseudopotential implementation of RT-TDDFT in the Qb@ll branch of the Qbox code\unskip~\cite{364506:8093956,364506:8092424,364506:9893475} , in which we have implemented the real-time propagation of maximally localized Wannier functions. Details of the on-the-fly localization algorithm and its performance and accuracy are given in the subsections that follow.

\subsubsection{Maximal Localization Procedure}The method employed in this work for computing MLWFs and time-dependent MLWFs (TD-MLWFs) is an extension of the method developed by Gygi, et al. \unskip~\cite{364506:8093500}. Gygi, et al. developed a parallelized Cardoso-Souloumiac diagonalization algorithm for the efficient and accurate computation of MLWFs in molecular and extended systems.  While this implementation was restricted to real wave functions (i.e., static ground state wave functions), the algorithm itself, as suggested by the authors, could be extended to complex wave functions (i.e., time-dependent wave functions), and in this work we have carried out such an extension.  What follows is a brief outline of the MLWF algorithm generalized for complex wave functions. Further details can be found in the works by Gygi, et al. \unskip~\cite{364506:8093500}.

In seminal work by Resta \unskip~\cite{364506:8092421}, the quantum-mechanical position operator is defined for periodic systems and its respective spread functional. In a rectangular periodic cell of dimensions $L_x,\;L_y,\;L_z $, at the center of the Brillouin zone, it can be shown that the spread associated with the position operator is equal to using the spread functional associated with a set of six self-adjoint operators $\left\{\widehat A^{(k)}\right\},\;k=1,\dots,6 $, defined as follows.  For an N-electron system, and the set of KS (or TDKS) orbitals $\left\{\phi_i\right\},\;i\;=\;1,\dots,\;N, $
\begin{eqnarray*}\sigma _{\left\{\widehat A^{(k)}\right\}}^{2}\;\left(\left\{\phi_i\right\}\right)\equiv\sum_{k=1}^{6}\sum_i\;\left[\left\langle\phi_i\left|\widehat A^{2}\right|\phi_i\right\rangle-\left\langle\phi_i\left|\widehat A\right|\phi_i\right\rangle^{2}\right]\end{eqnarray*}

\let\saveeqnno\theequation
\let\savefrac\frac
\def\dispfrac{\displaystyle\savefrac}
\begin{eqnarray}
\let\frac\dispfrac
\gdef\theequation{23}
\let\theHequation\theequation
\label{disp-formula-group-49dc00b5a337833868e25f912d87190a}
\begin{array}{@{}l}\begin{array}{rcl}&&\begin{array}{l}\widehat A^{(1)}\equiv\cos\frac{2\pi}{L_x}x\\\\\widehat A^{(2)}\equiv\sin\frac{2\pi}{L_x}x\\\\\widehat A^{(3)}\equiv\cos\frac{2\pi}{L_y}y\\\\\widehat A^{(4)}\equiv\sin\frac{2\pi}{L_y}y\\\\\widehat A^{(5)}\equiv\cos\frac{2\pi}{L_z}z\\\\\widehat A^{(6)}\equiv\sin\frac{2\pi}{L_z}z\end{array}\\&&\end{array}\end{array}
\end{eqnarray}
\global\let\theequation\saveeqnno
\addtocounter{equation}{-1}\ignorespaces 
Minimization of $\sigma _{\left\{\widehat A^{(k)}\right\}}^{2}\;\left(\left\{\phi_i\right\}\right) $ is achieved by calculating a unitary transformation matrix $U $ that simultaneously maximally diagonalizes the matrices $\widehat A^{(k)} $. This simultaneous diagonalization is achieved via the Cardoso-Souloumiac algorithm \unskip~\cite{364506:8093501}.  The diagonalization is carried out iteratively until the spread functional $\sigma _{\left\{\widehat A^{(k)}\right\}}^{2}\; $ converges to a minimum within a chosen tolerance. The resultant diagonal elements of the matrices $\widehat A^{(k)} $ can be used to determine the positions of the MLWF centers (MLWFCs) and the spreads of the MLWFs.  The iterative implementation of the Cardoso-Souloumiac algorithm for the computation of this transformation matrix $U $ is further detailed in the work by Gygi, et al.\unskip~\cite{364506:8093500}, with the differences for the case of real matrices noted. 

Because the calculation of this unitary MLWF transformation (i.e., unitary localization operator) is carried out at each RT-TDDFT step (sometimes for thousands of simulation steps), computational performance is of great importance. Like the Jacobi algorithm, the Cardoso-Souloumiac algorithm is inherently parallel. The current implementation makes use of MPI and high-performance linear algebra libraries (ScaLAPACK, BLACS) in conjunction with a processor-data rotation scheme \unskip~\cite{364506:8113631} for the efficient parallelized computation of the MLWF transformation.  Scaling and performance data for the TD-MLWF implementation compared with the standard TDKS implementation of RT-TDDFT are shown in the next section.

\subsubsection{Scaling and Accuracy}The calculation and application of the TD-MLWF unitary transformation during the RT-TDDFT propagation adds additional computational cost in the simulations stemming from the calculation of the localization operator. Thus, it is important to examine the performance and accuracy of the RT-TDDFT/TD-MLWF implementation relative to the standard RT-TDDFT/TDKS approach.  As test systems, we chose two representative cases: an isolated benzene molecule and crystalline silicon.  All of these simulations used the LDA exchange-correlation functional and Hamann-Schluter-Chiang-Vanderbilt (HSCV) norm-conserving pseudopotentials \unskip~\cite{364506:8093506}.  The adiabatic approximation was used for the time dependence of the exchange correlation functional \unskip~\cite{364506:8860372,364506:8860414}. First, standard DFT calculations were performed to acquire the ground state wave functions. Then, in order to create a perturbed non-equilibrium initial condition for the RT-TDDFT simulations, the atoms were translated by +0.01 Bohr in the x-direction at the start of the time-propagation.  For each system, we performed two simulations: a control case in which the TDKS equations are propagated with TDKS Bloch orbitals, and a test case involving TD-MLWF propagation. For all TD-MLWF simulations, a convergence tolerance of 10\ensuremath{^{-8}} was used in the joint approximate diagonilization algorithm described in the previous section.

The scalability tests for the benzene case were performed using a 30 Rydberg planewave kinetic cutoff energy, a 50 x 50 x 50 Bohr simulation cell, a 0.1 a.u. time step, for 100 simulation steps with the enforced time-reversal symmetry (ETRS) integrator \unskip~\cite{364506:8093645}.  The simulations were performed using MPI on nodes with 44 Intel Xeon processors each. Figure~\ref{f-33edac16ff73} shows the results of the performance test, and for this small case, the TD-MLWF propagation approach does not add any appreciable computational cost until we approach several hundred cores.  On 352 cores, a single TD-MLWF simulation step is 1.41 times the cost of a single TDKS simulation step. However, we also observe that the performance of the standard TDKS simulation also decreases when we reach over 500 cores. For such a small system (30 electrons), this is not surprising that we reach this performance bottleneck at only a few hundred cores.  

For testing scalability with the silicon crystal, a 512-atom supercell was used, with a 60 Rydberg planewave kinetic cutoff energy, a 0.1 a.u. time step, for 100 simulation steps with the ETRS integrator.  The simulations were performed using MPI on an IBM BG/Q system with 16 MPI tasks per node.  The results, shown inFigure~\ref{f-c9a1669663a1}, show that in all cases studied the TD-MLWF propagation comes at a greater computational cost than TDKS propagation, as expected. However, scaling up to larger numbers of cores, the performance gap slightly decreases between the two methods: On 256 cores, each TD-MLWF step is 1.88 times the cost, and on 32,768 cores, each TD-MLWF step is 1.88 times the cost.  The TD-MLWF propagation simulations in these test simulations used a very strict convergence criteria for the calculation of the localization operator, and with relaxation of this criteria the computational cost of the TD-MLWF propagation could be reduced even further while maintaining good accuracy. 

The scalability results in this work show that the calculation and propagation of TD-MLWFs incurs an additional computational cost, but that the additional cost does not significantly impede practical simulations. That being said, we have not yet explored the possibilities for TD-MLWFs to be used to actually reduce the cost of RT-TDDFT simulations in certain contexts.  To our knowledge, this work represents the first use of MLWF propagation in RT-TDDFT, but the spatially localized characteristics of MLWFs have been exploited in the context of ground-state DFT calculations in insulating systems to achieve order-N calculations of exact exchange \unskip~\cite{364506:8842590} as well as order-N first-principles molecular dynamics simulations based on a "divide and conquer" scheme \unskip~\cite{364506:8842632}.  The efficient calculation of hybrid exchange could allow for RT-TDDFT simulations with hybrid exchange-correlation functionals.  Additionally, the "divide and conquer" method described by Osei-Kuffour, et al. \unskip~\cite{364506:8842632} could potentially be used in large-scale RT-TDDFT simulations to improve performance by avoiding global communications and possibly by limiting the number of orbitals that need to be propagated at each time step. We plan to investigate these possibilities in the future.

In addition to testing the scalability of our TD-MLWF implementation, we performed simulations to test its accuracy.  In principle, when one transforms Bloch orbitals into MLWFs, all physical observables should remain unchanged because the transformation is unitary, preserving the total electron density.  However, this assumes that the transformation matrix is exactly unitary. Numerically, however, the joint diagonalization algorithm used to compute the matrix at each time step is approximate and requires us to choose a convergence tolerance for the spread functional (see "Maximal Localization Procedure" section).  Thus, it is important to compare the TD-MLWF and standard TDKS (Bloch orbitals) RT-TDDFT simulations to ensure that we can achieve excellent agreement in physical observables such as total electronic energy.  Again, we chose two representative systems as test cases: an isolated benzene molecule in a 50 x 50 x 50 Bohr simulation cell as well as a 64-atom crystalline silicon supercell.  In order to examine the accuracy of TD-MLWF propagation with different numerical integrators, we performed simulations with both the fourth-order Runge-Kutta (FORK) method \unskip~\cite{364506:8842633,364506:8093956} and the ETRS method \unskip~\cite{364506:8093645}.  As before, we initialize the RT-TDDFT simulations by shifting all atomic coordinates by +0.01 Bohr in the x direction relative to the coordinates used in the ground state DFT calculation. From this non-equilibrium starting point, the TDKS equations are propagated for 100 a.u. of time in two different ways: a control case in which the TDKS equations are propagated via Bloch orbitals, and a test case in which the TDKS equations are propagated in terms of the TD-MLWFs. With all atoms frozen, the total energy serves as a constant of motion, meaning that any drift in the energy can be attributed to numerical errors in the real-time propagation. For the TD-MLWF cases, a convergence tolerance of 10\ensuremath{^{-8}} was used for the joint approximate diagonilization algorithm. 

The results (see Figure~\ref{f-c9a1669663a1}) indicate that there is significant dependence on the numerical integrator used (FORK vs. ETRS), with the ETRS integrator conserving the total energy in the system better than FORK for a given $\Delta t $, especially for the case of the benzene molecule. Also, the ETRS propagator can use much larger time steps than the FORK (0.25 a.u. vs. 0.15 a.u.) before becoming completely unstable.  However, in al cases there is no significant difference (\textless\ 0.1\%) between the RT-TDDFT simulations with the standard Bloch TDKS propagation and the TD-MLWF propagation.  This shows that our TD-MLWF implementation is capable of maintaining high numerical accuracy in the computation and application of the unitary gauge transformation at every time step.
    
\section{Results and Discussion}

\subsection{Optical Excitation}

\subsubsection{Benzene Molecule}As an example application for the real-time propagation of MLWFs, we have performed RT-TDDFT simulations on a benzene molecule in vacuum to calculate the optical absorption spectrum. The optical properties of gas-phase benzene have been well-studied both in experimental and theoretical work.  In particular, the optical absorption spectrum has been calculated via both RT-TDDFT \unskip~\cite{364506:8113521,364506:8113522} and the Liouville-Lanczos approach to linear response TDDFT (LR-TDDFT) \unskip~\cite{364506:8298352}, and in the case of gaseous benzene, the calculated spectra are in very good agreement with reported experimental results, even with the LDA exchange-correlation approximation \unskip~\cite{364506:8093328,364506:8093327}.

In our RT-TDDFT approach, a static polarized state electronic structure was first obtained adiabatically via a DFT calculation with an electric field of magnitude 0.001 a.u. present using the MLWF/length-gauge formulation described in the Theoretical and Computational Methods section. This electric field magnitude is sufficiently small to ensure a linear response, and to ensure oscillations of the electronic dipole that are large enough compared to any numerical noise. The LDA exchange-correlation (XC) functional was used in all simulations. The basis was expanded in planewaves up to a kinetic cutoff energy of 30 Ryd. A large cubic simulation cell of 100 x 100 x 100 Bohr was used, with the benzene ring lying in the xy plane. Such a large simulation cell was used in order to avoid interactions between periodic images of the system which can lead to extra "ripples" in the absorption spectrum for high-energy excitations where ionization becomes important \unskip~\cite{364506:8093509}. Another approach to mitigate such finite size effects is to employ a complex absorbing potential at the simulation cell boundaries \unskip~\cite{364506:8093509}, which is a capability that we plan to implement in the future. The RT-TDDFT simulations were initialized from the static field-polarized MLWFs. The enforced time reversal symmetry (ETRS) \unskip~\cite{364506:8093645} was used for the numerical propagation of the TDKS equations.  A time step of 0.1 a.u. was used. At the initial RT-TDDFT time step, the electric field is "turned off", and the system propagates for 500 a.u. for each simulation. This sudden "switching off" of the electric field introduces a phase to the wavefunctions, resulting in continuous oscillations of the electronic dipole.  Throughout the RT-TDDFT simulations, nuclear motion would be negligible due to the relatively short timescales and the small perturbations, thus, the nuclear positions are held fixed in order to reduce computational expense.

\bgroup
\fixFloatSize{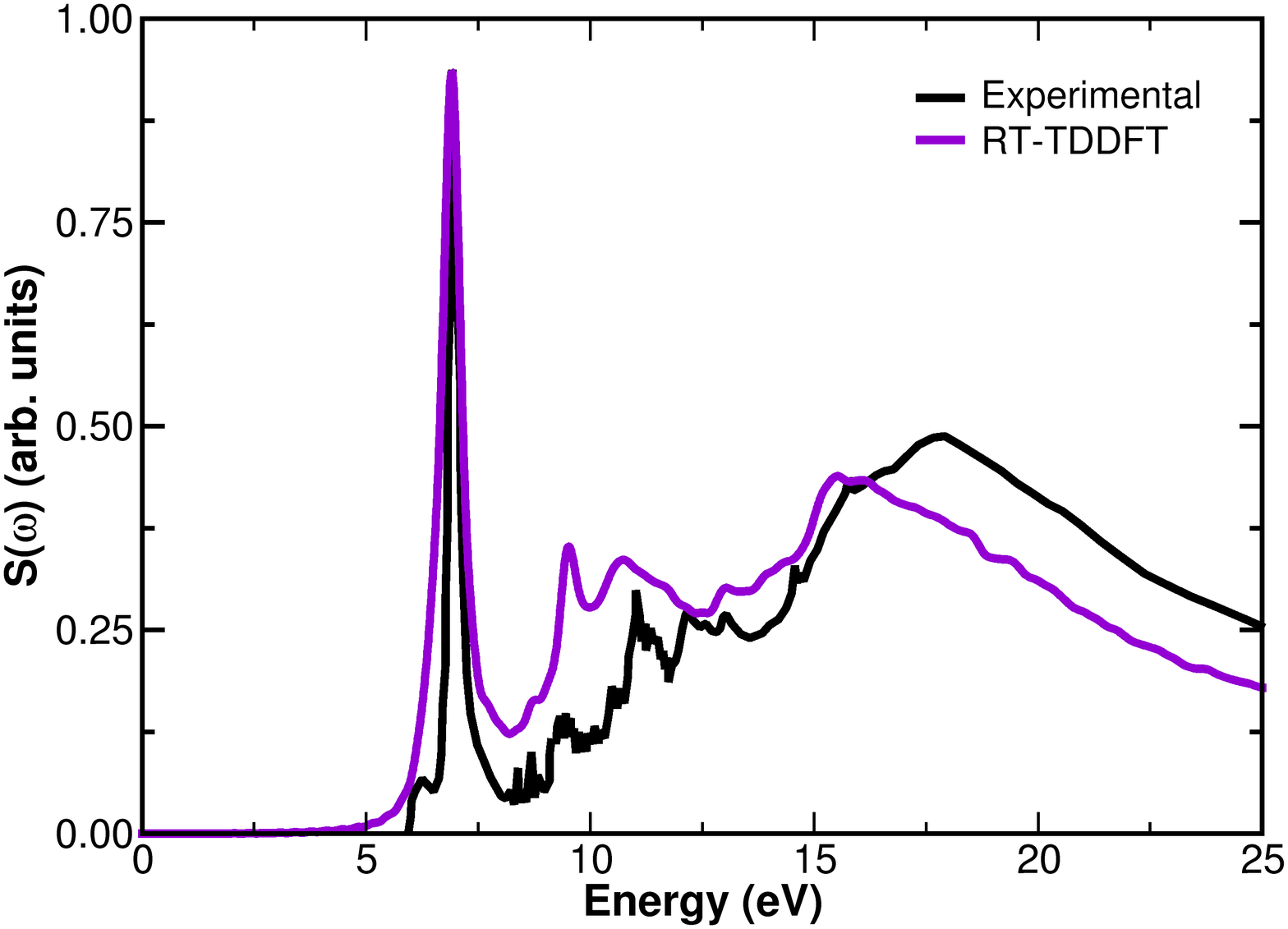}
\begin{figure}[t!]
\centering \makeatletter\IfFileExists{images/a22409bd-ff82-456b-a973-f09703d6a730-ufig4.png}{\includegraphics{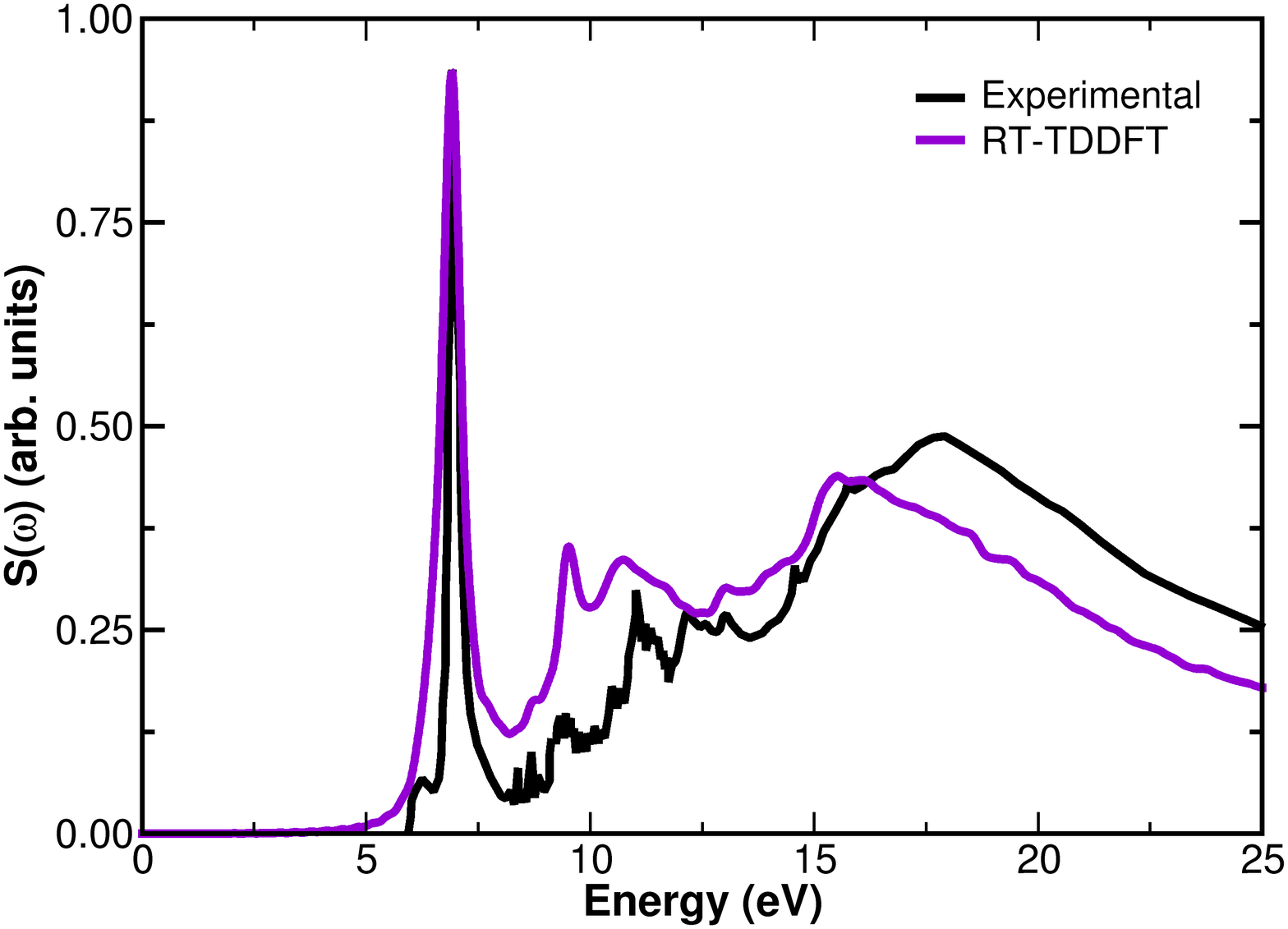}}{}
\makeatother 
\caption{{Dipole strength function (i.e., electronic absorption spectrum) for gas phase benzene calculated via TD-MLWFs/RT-TDDFT in this work (purple) compared to experimental data (black). }}
\label{f-da66}
\end{figure}
\egroup
As can be seen in Figure~\ref{f-da66}, the RT-TDDFT calculated spectrum shows good agreement with the experimental data.  Indeed, other studies have shown that the optical absorption spectrum of molecular benzene can be successfully determined with TDDFT calculations \unskip~\cite{364506:8113521}, even with the simple LDA approximation to the exchange-correlation functional.

\bgroup
\fixFloatSize{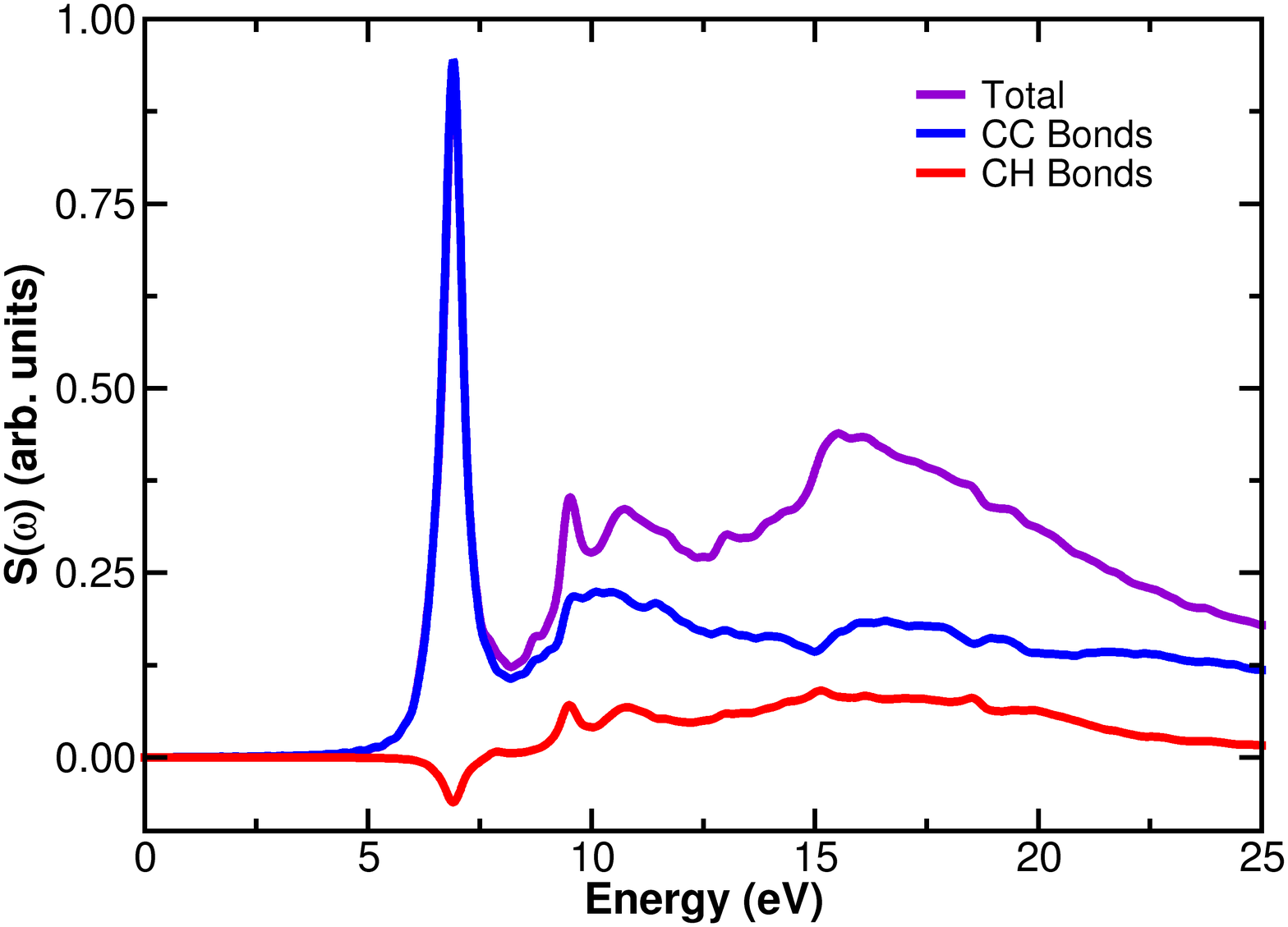}
\begin{figure}[t!]
\centering \makeatletter\IfFileExists{images/eca10ba6-ce72-4c00-9202-404472d5836c-ufig5.png}{\includegraphics{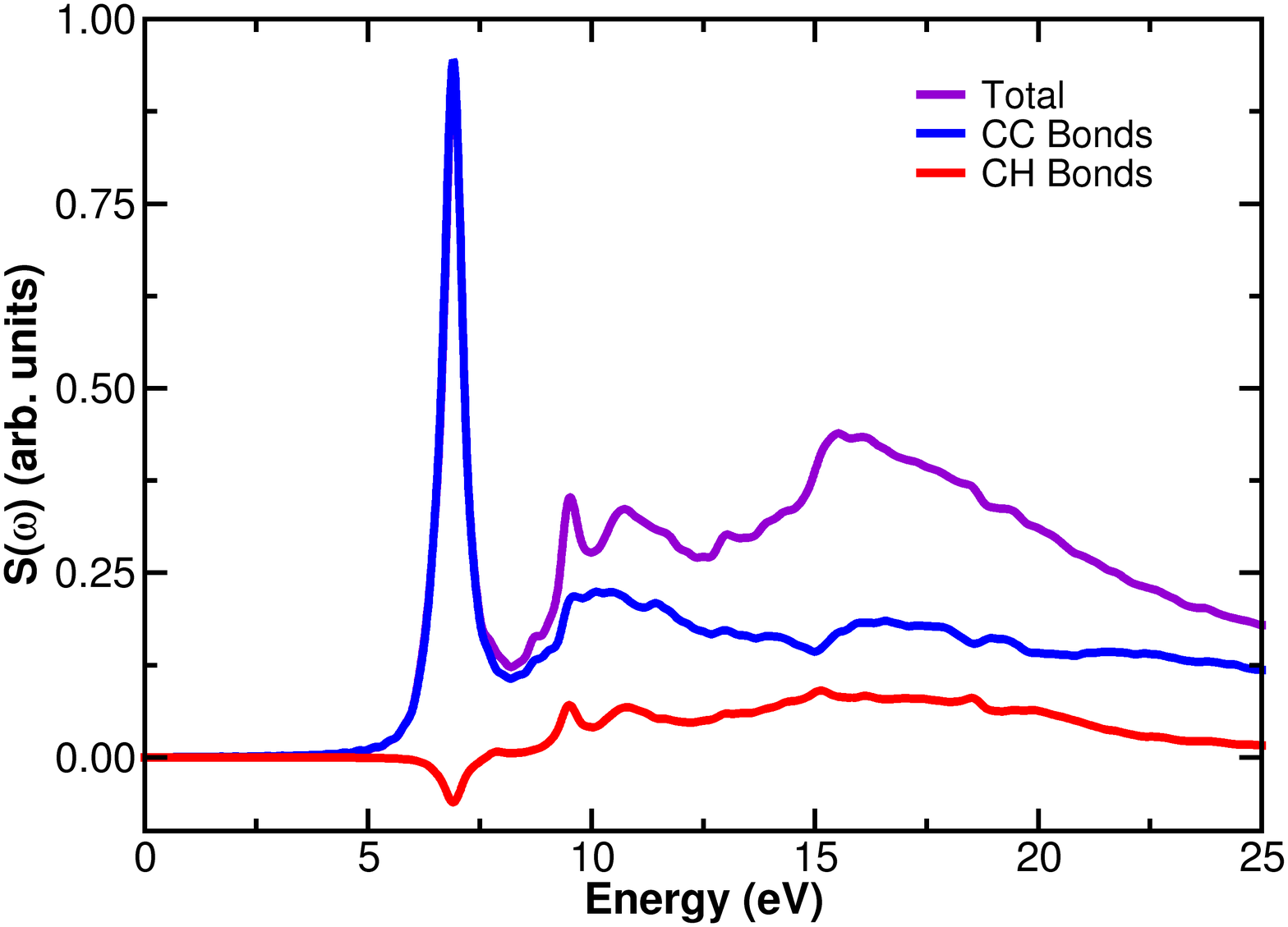}}{}
\makeatother 
\caption{{Dipole strength function for gas-phase benzene (purple) decomposed via TD-MLWF centers into carbon-carbon bonds (blue) and carbon-hydrogen bonds (red). The negative values observed in the CH bond spectrum indicate oscillations that are out of phase with the CC bond oscillations. }}
\label{f-4c84}
\end{figure}
\egroup
In addition to calculating the photo-absorption spectrum of benzene, the TD-MLWF approach allows us to decompose the spectrum into contributions from different chemical moieties.  For molecular systems, the maximal localization transformation results in Wannier functions that are localized on bonds, and/or lone pairs. Thus, in the case of benzene, the TD-MLWFs and their corresponding centers can easily be associated with either carbon-carbon  (CC) or carbon-hydrogen (CH) bonds. By Fourier-transforming the dynamic polarization of just the CC bonds or just the CH bonds, we can acquire the chemically-decomposed spectra shown in Figure~\ref{f-4c84}.  Interestingly, there are portions of the C-H spectrum which are negative. While negative values for the total absorption would be unphysical, the decomposed spectrum allows for this possibility.  The negative "absorption" indicates that, at {\texttildeapprox}7 eV, the oscillations of the CH bonds are out of phase with those of the CC bonds. This destructive interference results in total dynamic polarization oscillations that are smaller in magnitude than the CC bond dipole oscillations themselves.

\subsubsection{Water Molecule}
\bgroup
\fixFloatSize{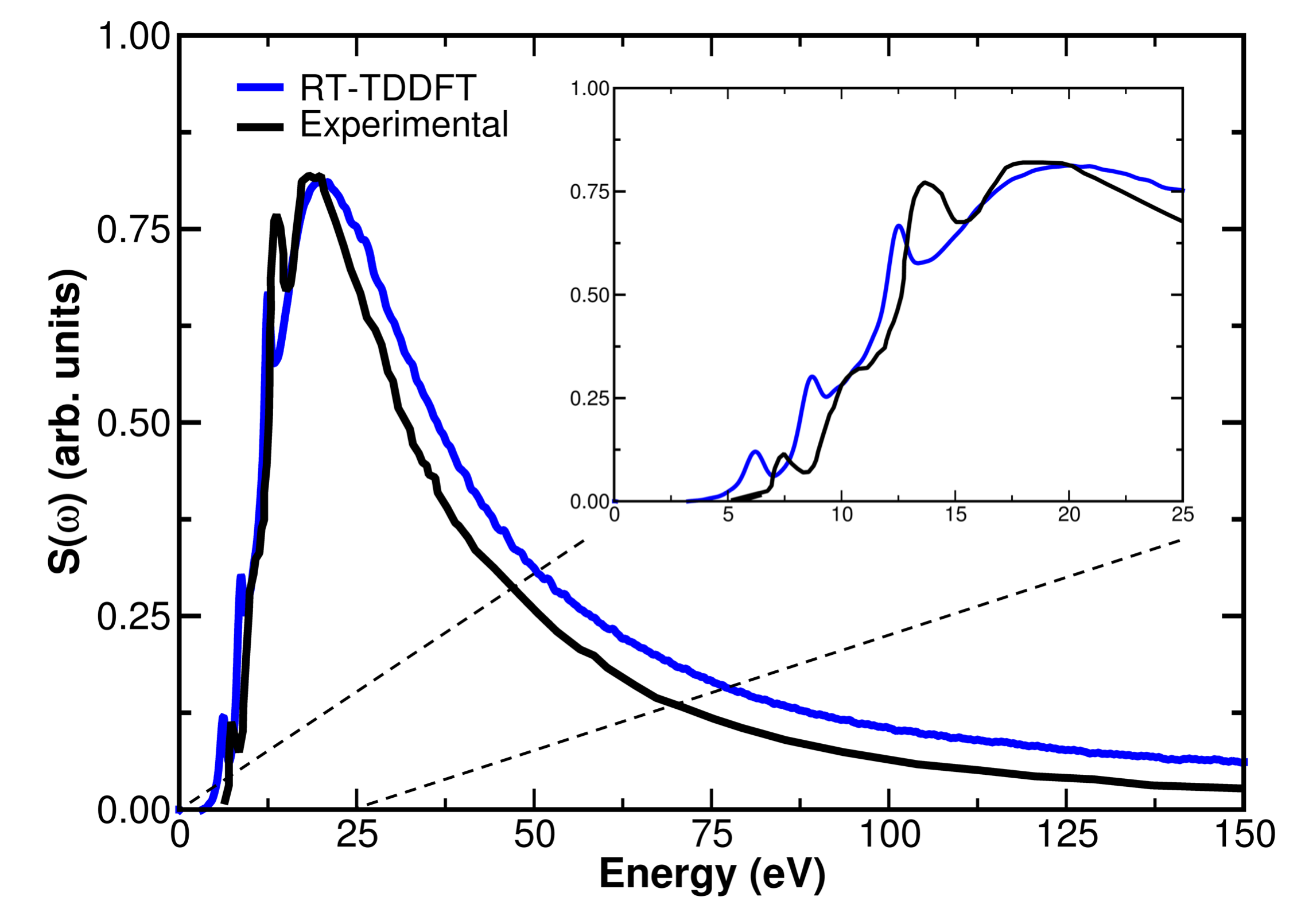}
\begin{figure}[b!]
\centering \makeatletter\IfFileExists{images/2001dfcc-4ad4-4f06-9f4d-e78764119d2d-ufig6.png}{\includegraphics{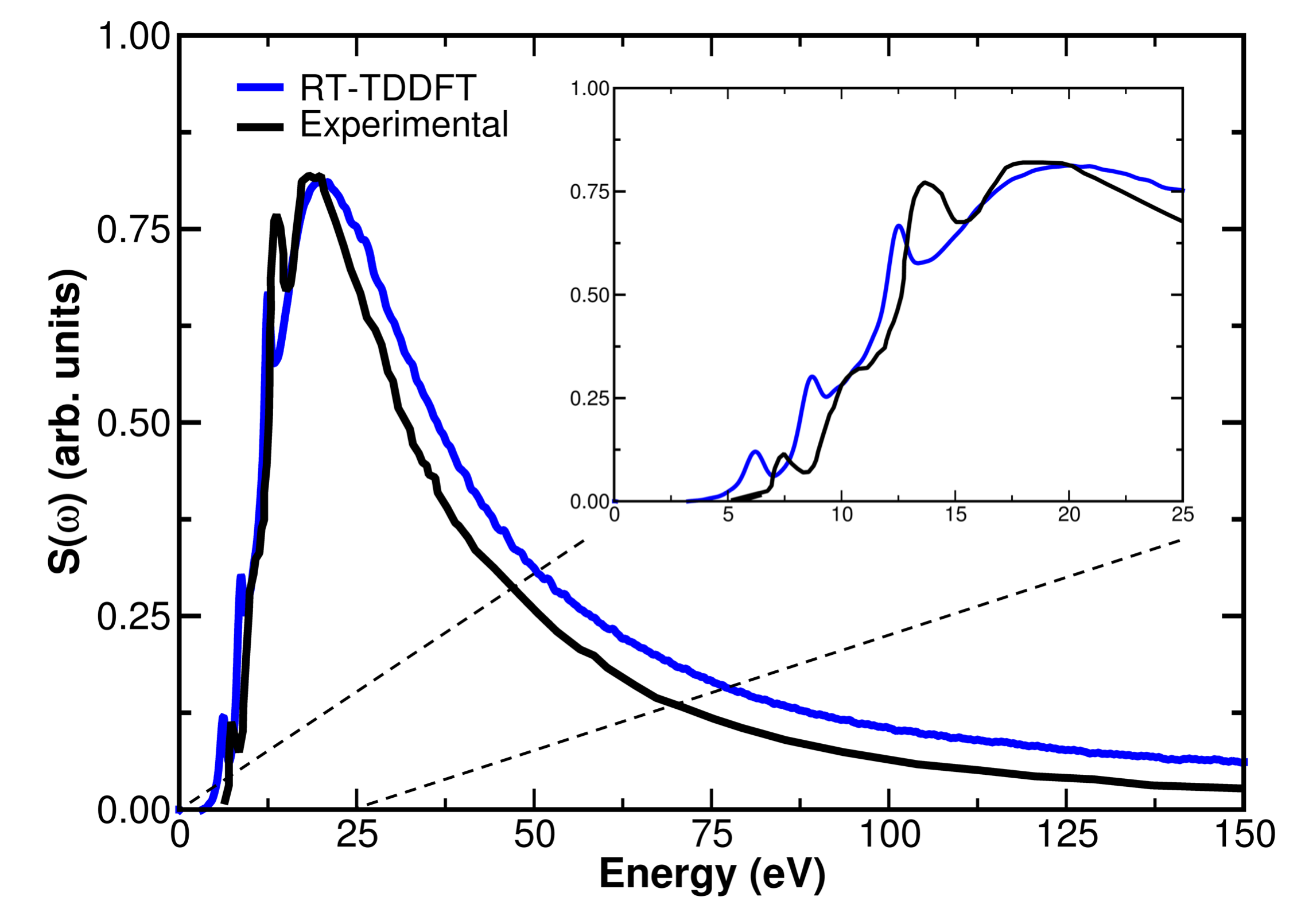}}{}
\makeatother 
\caption{{Dipole strength function (i.e., electronic absorption spectrum) for gas phase water calculated via TD-MLWFs/RT-TDDFT in this work (blue) compared to the experimental spectrum (black). The inset more clearly shows the low-energy portions of the spectra. }}
\label{figure-198fffb19319295c635a178737564090}
\end{figure}
\egroup
Due to the availability of experimental absorption spectrum data over a wide range ({\texttildeapprox}0 eV to {\texttildeapprox}200 eV), gaseous water is another interesting test case.  LR-TDDFT based on Casida's equation \unskip~\cite{364506:8296709} is the most widely used approach for computing absorption spectra of materials and molecules.  However, Casida's equation is solved iteratively in a basis of (occupied) x (virtual) dimensions \unskip~\cite{364506:8297122}, meaning that calculations involving more than a few low-lying excited states become computationally prohibitive.  Thus, the calculation of the broad-band spectrum of gaseous water is one problem that lends itself to RT-TDDFT. Additionally, RT-TDDFT methods in planewave bases have the capabilities to capture high-energy excitations. This is unlike atom-centered basis set RT-TDDFT methods, which are known to exhibit spurious high-energy artifacts unless unlike some localized basis implementations which can erroneously predict spurious excitations at high energies unless an imaginary molecular orbital-based absorbing potential is used \unskip~\cite{364506:9399471}.

The simulation details closely follow those described for the benzene molecule case. A static polarized state electronic structure was first determined adiabatically via a DFT calculation with an electric field of magnitude 0.001 a.u. present. The PBE exchange-correlation functional\unskip~\cite{364506:8093504} was used in all simulations. The basis was expanded in planewaves up to a kinetic cutoff energy of 40 Ryd. A cubic simulation cell of 100 x 100 x 100 Bohr was used.  The enforced time reversal symmetry (ETRS) \unskip~\cite{364506:8093645} was used for the numerical propagation of the TDKS equations.  A time step of 0.05 a.u. was used, with a total propagation of 250 a.u. for each simulation. At the initial RT-TDDFT time step, the system is perturbed by suddenly switching off the static homogeneous electric field. This perturbation introduces a phase to the wavefunctions, resulting in continuous oscillations of the electronic dipole. This electric field magnitude is sufficiently small to ensure a linear response, to conserve energy throughout the numerical propagation, and to acquire oscillations of the electronic dipole that are large compared to any numerical noise. The nuclei are held in fixed positions throughout the RT-TDDFT simulations.

The RT-TDDFT calculated spectrum alongside the experimental spectrum \unskip~\cite{364506:8864196} are shown in Figure~\ref{figure-198fffb19319295c635a178737564090} in two different energy ranges, with the absorption spectra in a lower energy range (0 eV to 30 eV), displaying three well-defined peaks in addition to the rapid onset of broadband absorption. While the RT-TDDFT results show good qualitative agreement with the experimental spectrum, there is a red-shift of the peaks by {\texttildeapprox}2 eV.  It is possible that RT-TDDFT with hybrid XC functionals or meta-GGA functionals could yield spectra in better agreement to experiment, and we plan to explore this in a future work. We also see that the broadband spectrum (0 eV to 150 eV) is in overall agreement with the experimental data.  This showcases the ability of the abilities of the RT-TDDFT simulations to acquire the absorption spectrum over a wide energy range, including photoemission-like excitations.

\bgroup
\fixFloatSize{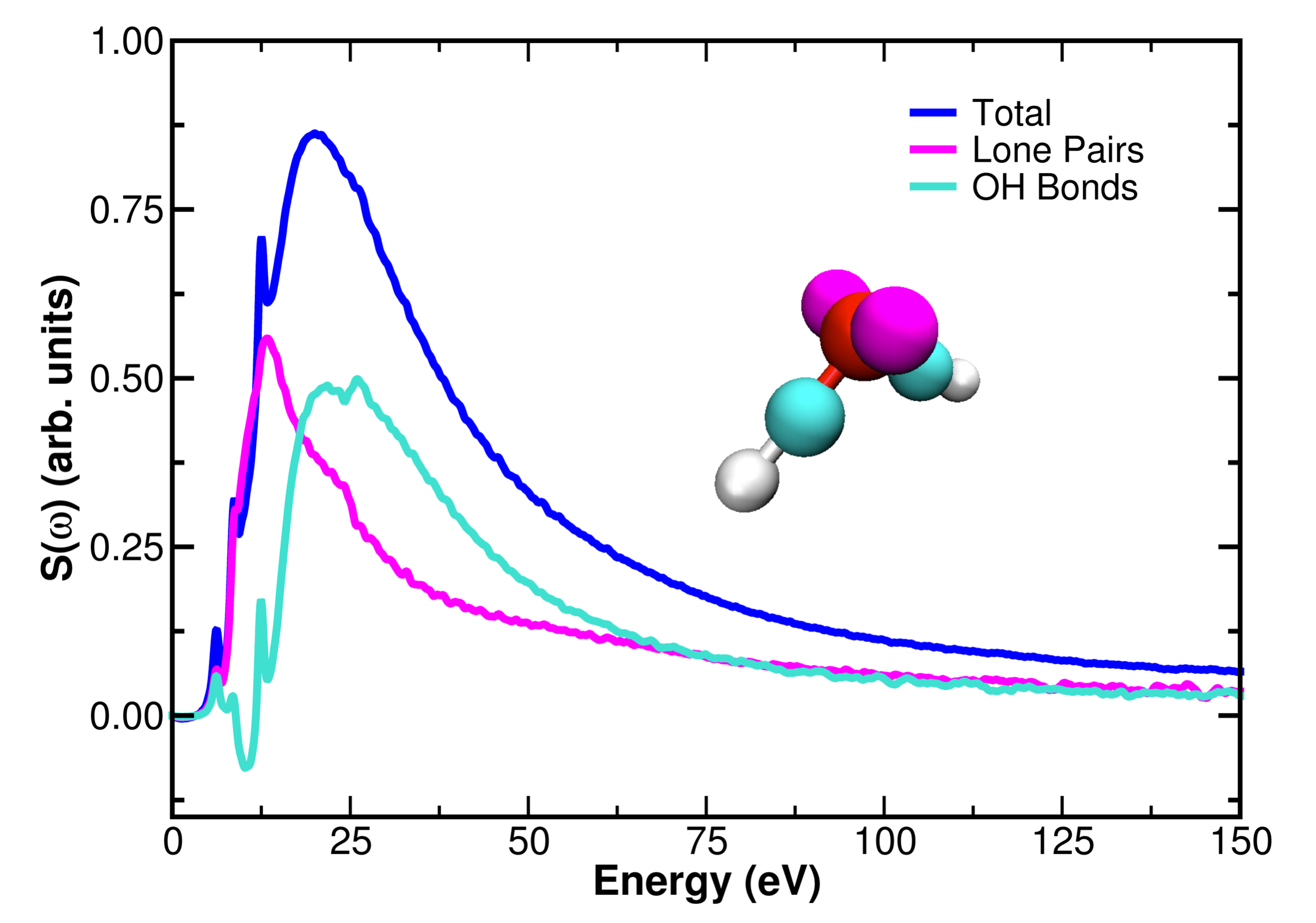}
\begin{figure}[t!]
\centering \makeatletter\IfFileExists{images/eadb17df-1c54-4d25-bf7d-309875648008-ufig7.png}{\includegraphics{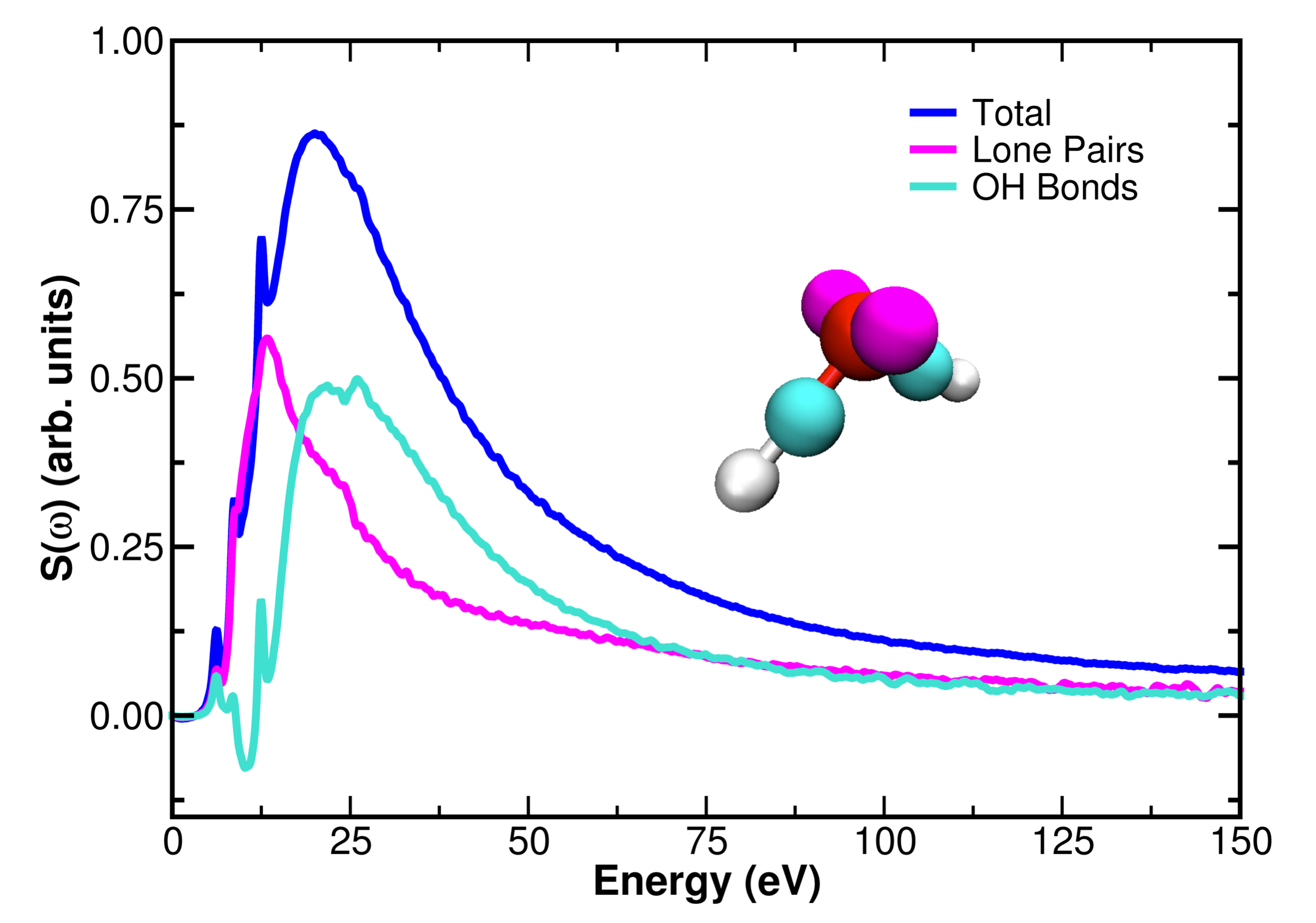}}{}
\makeatother 
\caption{{Dipole strength function for gas-phase water (blue) decomposed via TD-MLWF centers into oxygen-hydrogen bonds (light blue) and lone pairs (red). The negative values observed in the OH bond spectrum indicate oscillations that are out of phase with the lone pair bond oscillations. }}
\label{f-b193}
\end{figure}
\egroup
Similar to the molecular benzene case, the TD-MLWFs allow us to decompose the spectrum for a more detailed chemical understanding of the photoabsorption trends.  The spectrum, decomposed in terms of contributions from oxygen-hydrogen (OH) bond TD-MLWFs and lone pair (LP) TD-MLWFs is shown in Figure~\ref{f-b193}. For the OH bond spectrum, we again see a region of negative magnitude, implying destructive interference in the coupling between OH bonds and LPs. Also, we observe that the LP spectrum does not contain sharp peaks as seen in the OH bond spectrum, which comprises the three well-defined peaks in the 0 to 15 eV range.  However, with regards to overall magnitude, the LP excitations dominate at low energies below {\texttildeapprox}15 eV, whereas the OH bond absorption dominates in a higher energy range.

\subsubsection{Benzene in Liquid Water}With the capability to calculated dynamic polarizability in both molecular and condensed matter systems, we can use the TD-MLWF approach to study complex systems such as molecules solvated in liquid water.  Known as the solvatochromic effect, the optical absorption spectrum of solute molecules, such as organic dyes, varies depending on the solvent \unskip~\cite{364506:8864279}.  Thus, it is often important to account for solvent effects when performing excited state calculations on molecules, materials, and nanostructures.  The most common approach to represent the solvent environment in TDDFT calculations is through a polarizable continuum model (PCM) \unskip~\cite{364506:8864280}, which is computationally inexpensive but may not accurately describe solvatochromic effects. Also, mixed quantum mechanical-classical (QM/MM) methods with standard or polarizable force fields can be used, providing a balance between accuracy and computational cost \unskip~\cite{364506:8864322,364506:8864323,364506:8092476}.  However, in QM/MM methods, partitioning QM region can be a somewhat arbitrary or ambiguous choice. Also, by construction, in QM/MM approaches the MM region cannot be photo-excited.  An alternative, albeit computationally expensive, approach is to treat the entire system (solute and solvent) quantum mechanically, performing TDDFT calculations on the full electronic structure \unskip~\cite{364506:8092477}.  With the TD-MLWF approach, we can propagate the full electronic system with the TDKS equations, with the additional benefit that the orbitals are localized, allowing for the calculation of dynamic polarization for any individual molecule, or for the liquid system as a whole.

\bgroup
\fixFloatSize{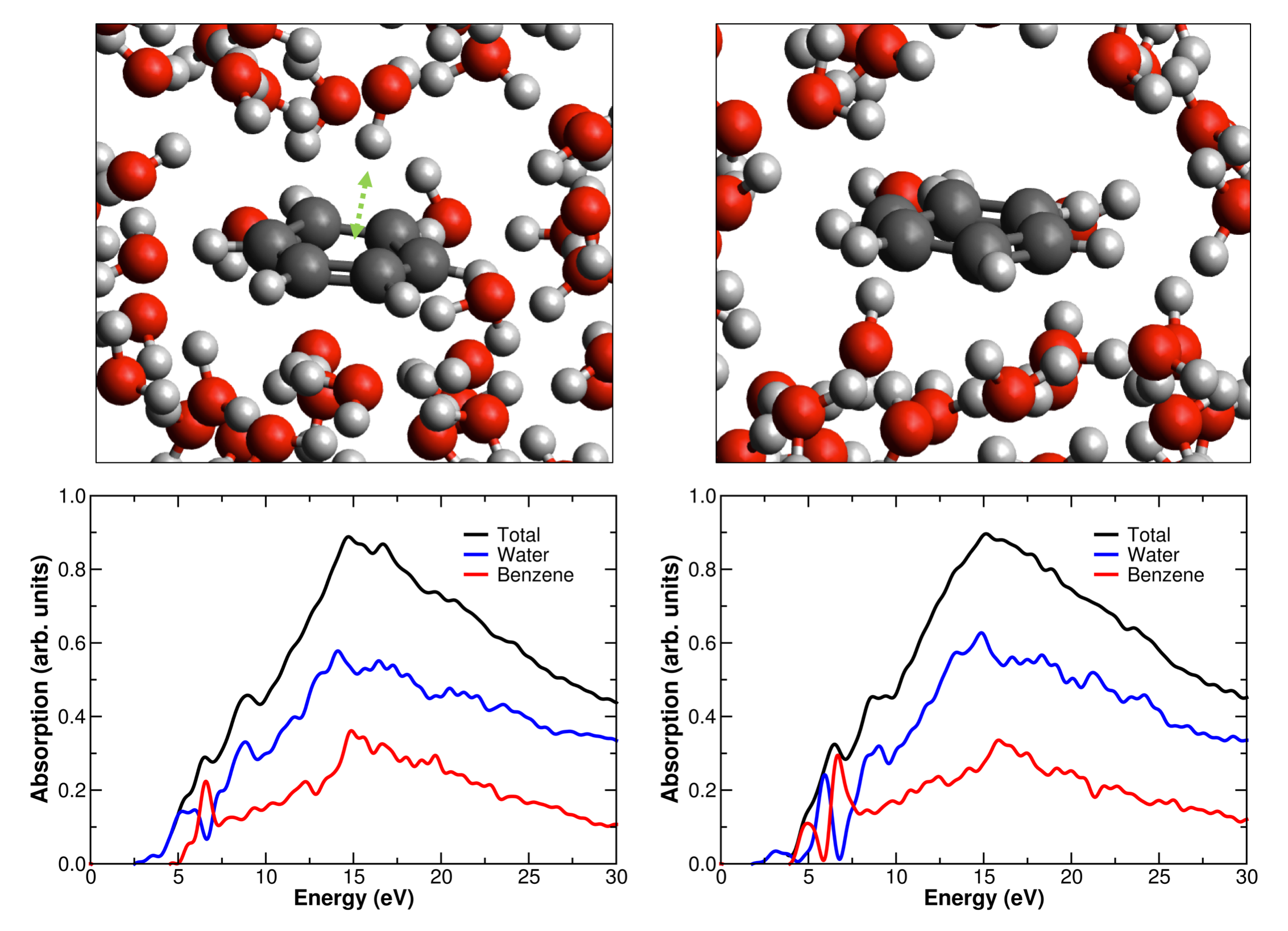}
\begin{figure*}[t!]
\centering \makeatletter\IfFileExists{images/46f19382-745d-46bf-9a29-4e98fd1c864a-ufig8.png}{\includegraphics[width=.69\linewidth]{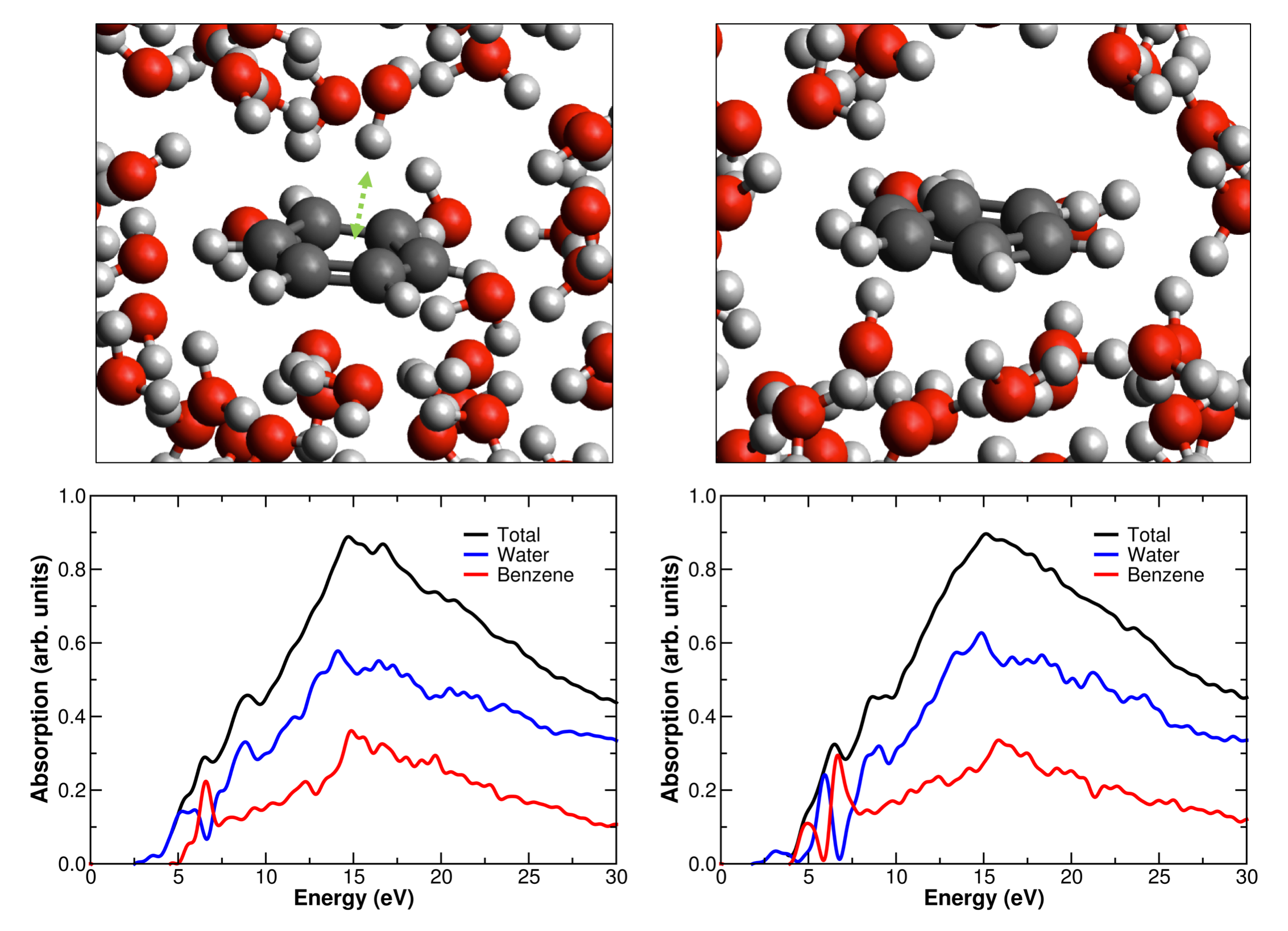}}{}
\makeatother 
\caption{{Absorption spectra for a benzene in water system (black) decomposed via TD-MLWF centers into contributions from the benzene molecule (red) and the water molecules (blue). Two configurations from two MD snapshots are represented: one structure with a water molecule hydrogen pointed into the benzene molecule pi cloud (left), and another without any such coordination (right). }}
\label{figure-29aca76e81821fe201e64468873a91f1}
\end{figure*}
\egroup
In order to demonstrate this approach, we again chose benzene as a test case. However, in this system, we explicitly solvated the benzene molecule in a cubic simulation cell (25.04 a.u. x 25.04 a.u. x 25.04 a.u.) containing 73 water molecules (614 electrons in total, including benzene). In order to acquire the absorption spectrum for a solvated molecule, one should in principle perform TDDFT calculations on a random ensemble of solvent/solute structures \unskip~\cite{364506:8864324}. However, due to the considerable computational cost of performing a RT-TDDFT calculation on even one structure containing over 600 electrons, we restrict ourselves here to two representative configurations known to be important in influencing the electronic structure of benzene \unskip~\cite{364506:8864325}, for the purposes of illustrating the TD-MLWF utility.  A first-principles molecular dynamics simulation using the CP2K code \unskip~\cite{364506:8298304} with the SCAN XC functional \unskip~\cite{364506:9153745} and the DZVP basis was used to acquire the two MD snapshots to use for the RT-TDDFT calculations. In Configuration 1, the hydrogen of a single water molecule is directed toward the center of the benzene ring (see Figure~\ref{figure-29aca76e81821fe201e64468873a91f1} top left). In Configuration 2, there is no such ``special'' benzene-water molecule interaction (see Figure~\ref{figure-29aca76e81821fe201e64468873a91f1} top right). For both configurations, a standard DFT calculation was performed to acquire the electronic ground states. Then, the RT-TDDFT simulations with TD-MLWF propagation were performed.  For all simulations, the PBE exchange-correlation functional was used.  Additional calculation parameters were: 0.05 a.u. time step, 250 a.u. total simulation time, 40 Rydberg planewave cutoff energy, 0.001 a.u. electric field impulse in x, y, and z directions (separately) at t = 0.

Figure~\ref{figure-29aca76e81821fe201e64468873a91f1} shows the calculated absorption spectra of the entire benzene and water system with the dynamic polarizability obtained from the RT-TDDFT simulations.  Using the positions of the TD-MLWF centers to decompose the dynamic polarization response, we also calculated the absorption spectrum of both the benzene molecule and liquid water separately. The total absorption spectrum does not have particularly well-separated peaks, but with this spectrum decomposition, the contributions from each molecular species becomes clearer. For the benzene molecule, there is a peak at {\texttildeapprox}7 eV, as was also seen in the vacuum environment. However, for configuration 2, there is an additional low-energy peak at {\texttildeapprox}5 eV.  These results illustrate the usefulness and flexibility of the TD-MLWF approach in uncovering molecular-level details of complex chemical phenomena such as the solvatochromic effect.

\subsubsection{Crystalline Silicon}
\bgroup
\fixFloatSize{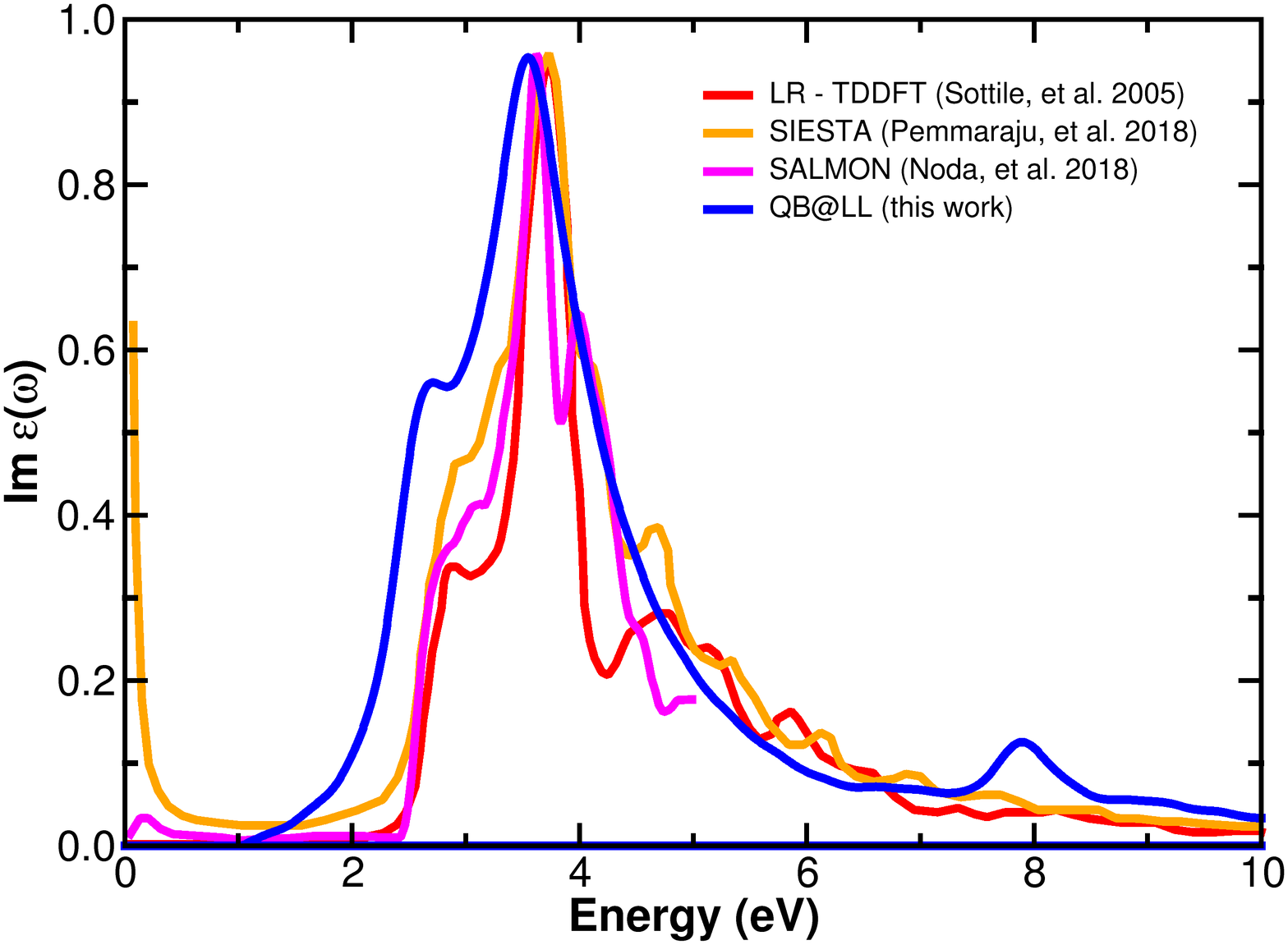}
\begin{figure}[t!]
\centering \makeatletter\IfFileExists{images/38b552c4-d7cc-4e21-a25f-9a4c8cf47fb4-ufig9.png}{\includegraphics{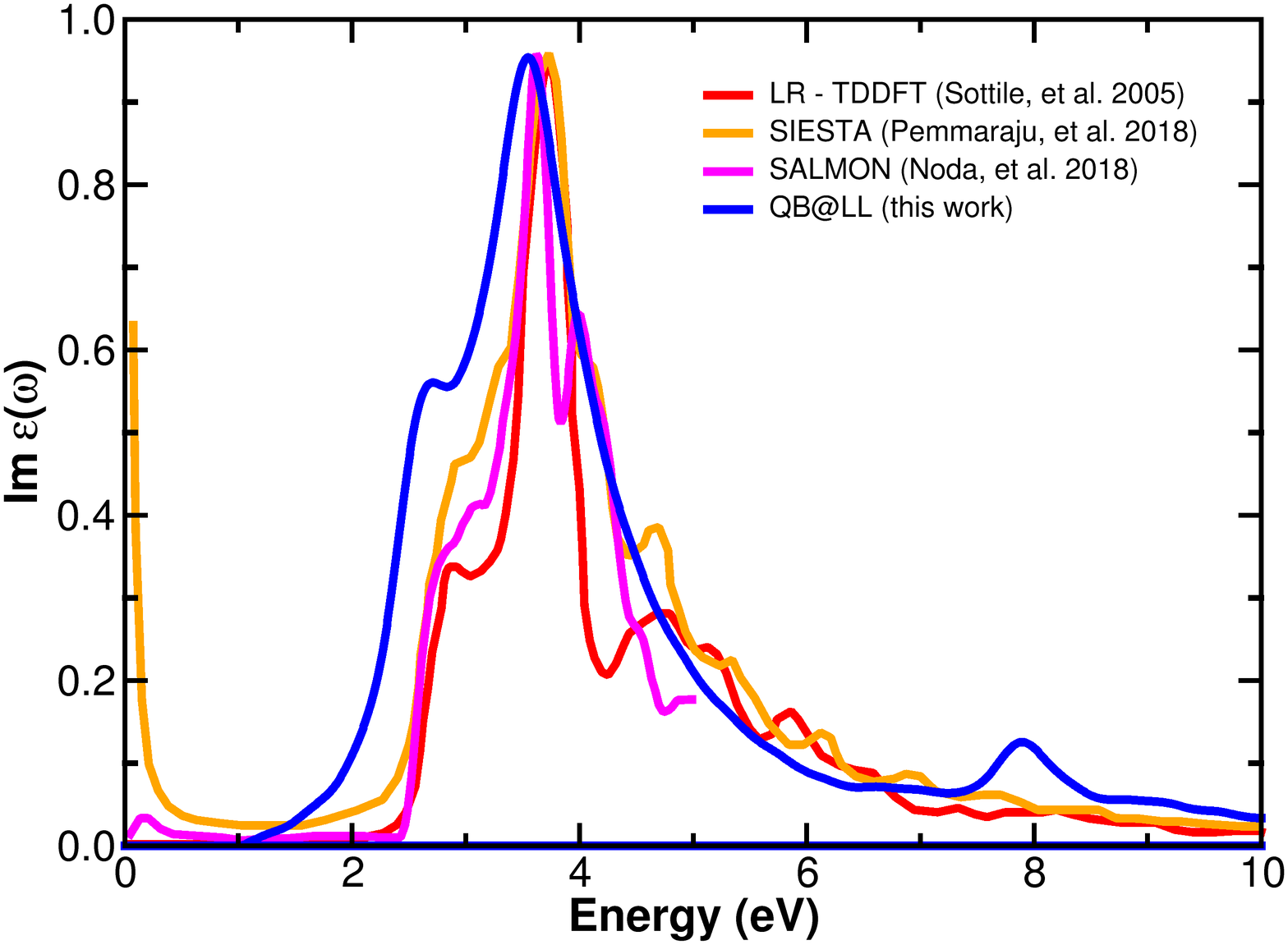}}{}
\makeatother 
\caption{{Imaginary part of the dielectric function for crystalline silicon calculated this work via TD-MLWFs/RT-TDDFT using the QB@LL code (blue). This spectrum is compared to reported results from LR-TDDFT (red), and two velocity-gauge RT-TDDFT calculations using the atomic orbital-based SIESTA code (orange) and the real-space-grid SALMON code (pink). }}
\label{figure-a38b621d761a18482fde8b350749b4d5}
\end{figure}
\egroup
As an example of the TD-MLWF approach for periodic systems, we have performed RT-TDDFT simulations on crystalline silicon due to the fact that this material has been studied by a range of RT-TDDFT codes, results which provide useful points of comparison. For all simulations, the LDA exchange correlation functional was used. A planewave kinetic energy cutoff of 40 Rydberg was used.  Additional calculation parameters are as follows: 0.05 a.u. time step, 250 a.u. total simulation time, 40 Rydberg planewave cutoff energy, 0.001 a.u. electric field impulse strength. 

For periodic crystalline systems one must take particular care to ensure convergence with respect to finite size effects. Our current TD-MLWF implementation does not yet include capabilities to incorporate multiple k-points, thus, in this work we use only the $\Gamma $ point and a silicon supercell in order to acquire accurate results for the bulk material.  Previous RT-TDDFT calculations on the optical properties of silicon have shown that k-point grids of 16x16x16 can be required to ensure full convergence of the absorption spectrum \unskip~\cite{364506:8092425}.  Because of this, converging the spectrum with respect to supercell size may seem like a computationally intractable problem, as it would require a simulation cell with tens of thousands of electrons. However, we can take advantage of the lattice symmetry to greatly reduce the cost of obtaining a converged absorption spectrum.  Due to the symmetry of the diamond cubic crystalline silicon, the linear response of the system will be identical for impulses in the x, y, and z directions.  Thus, the diagonal elements polarizability tensor (Equation~(\ref{disp-formula-group-05a7ec00aab60effb08e925c0363f14b})) will all be equal, implying that only one RT-TDDFT simulation is required to obtain the full linear response of the system. With regards to the dimensions of the silicon supercell, as shown in work by Darrigan, et al. \unskip~\cite{364506:8092478} one can achieve convergence with respect to cell dimensions simply by extending the supercell in the direction of the applied electric field.  Thus, if we are perturbing the system with an impulsive electric field in the x direction, we can simply repeat the 8-atom unit cell in the x direction until we reach convergence.  Figure S2 (supporting information) shows the resultant spectra for a range of supercells, and we observe that a 256-atom supercell is required to converge the imaginary part of the dielectric function.

In Figure~\ref{figure-a38b621d761a18482fde8b350749b4d5} we compare our results to those acquired via a range of other TDDFT codes.  The SALMON (scalable ab-initio light-matter simulator for optics and nanoscience) software package \unskip~\cite{364506:8298307} involves a RT-TDDFT implementation in which the time-dependent Kohn-Sham orbitals are discretized on a 3D real-space grid.  In the SALMON code, the perturbing field is represented in the velocity gauge by a shift of the vector potential, and the dielectric function is determined with respect to the induced electric current density \unskip~\cite{364506:8298307}.  The SIESTA code employs an RT-TDDFT calculations within a linear combination of atomic orbitals (LCAO) basis set framework, and the finite field calculations are also carried out in the velocity-gauge, with a vector potential representation of the perturbation \unskip~\cite{364506:8092425}.  Third, Figure~\ref{figure-a38b621d761a18482fde8b350749b4d5} includes results from linear-response TDDFT calculations in a planewave pseudopotential implementation \unskip~\cite{364506:8092474}. Although linear-response TDDFT represents a different theoretical framework altogether, the results should agree with those acquired through RT-TDDFT simulations, and, as can be seen in Figure~\ref{figure-a38b621d761a18482fde8b350749b4d5}, they do.  Generally speaking, all of the spectra are in good agreement, with the slight differences likely being attributable to numerical differences in the various implementations and basis sets. Although TDDFT is known to yield poor silicon spectra for conventional XC approximations, this example helps to illustrate the flexibility that real-time propagation of TD-MLWFs provides, eliminating any need to change gauge representations between isolated and periodic systems.

\subsection{Quantized Charge Transport}
\bgroup
\fixFloatSize{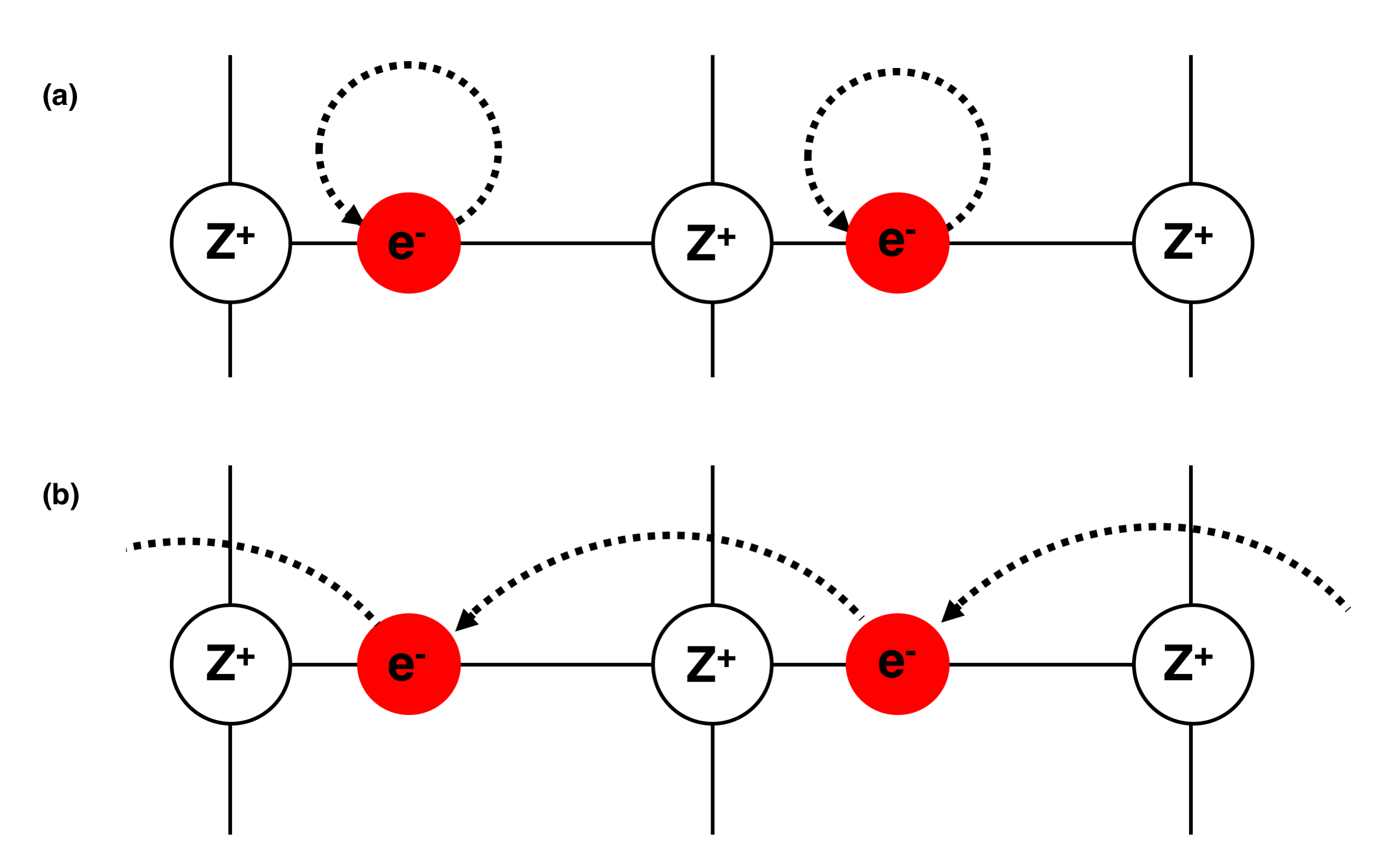}
\begin{figure}[t!]
\centering \makeatletter\IfFileExists{images/2b8d7948-d9d6-4dfe-9a8a-d9f4f96bd841-ufig10.png}{\includegraphics{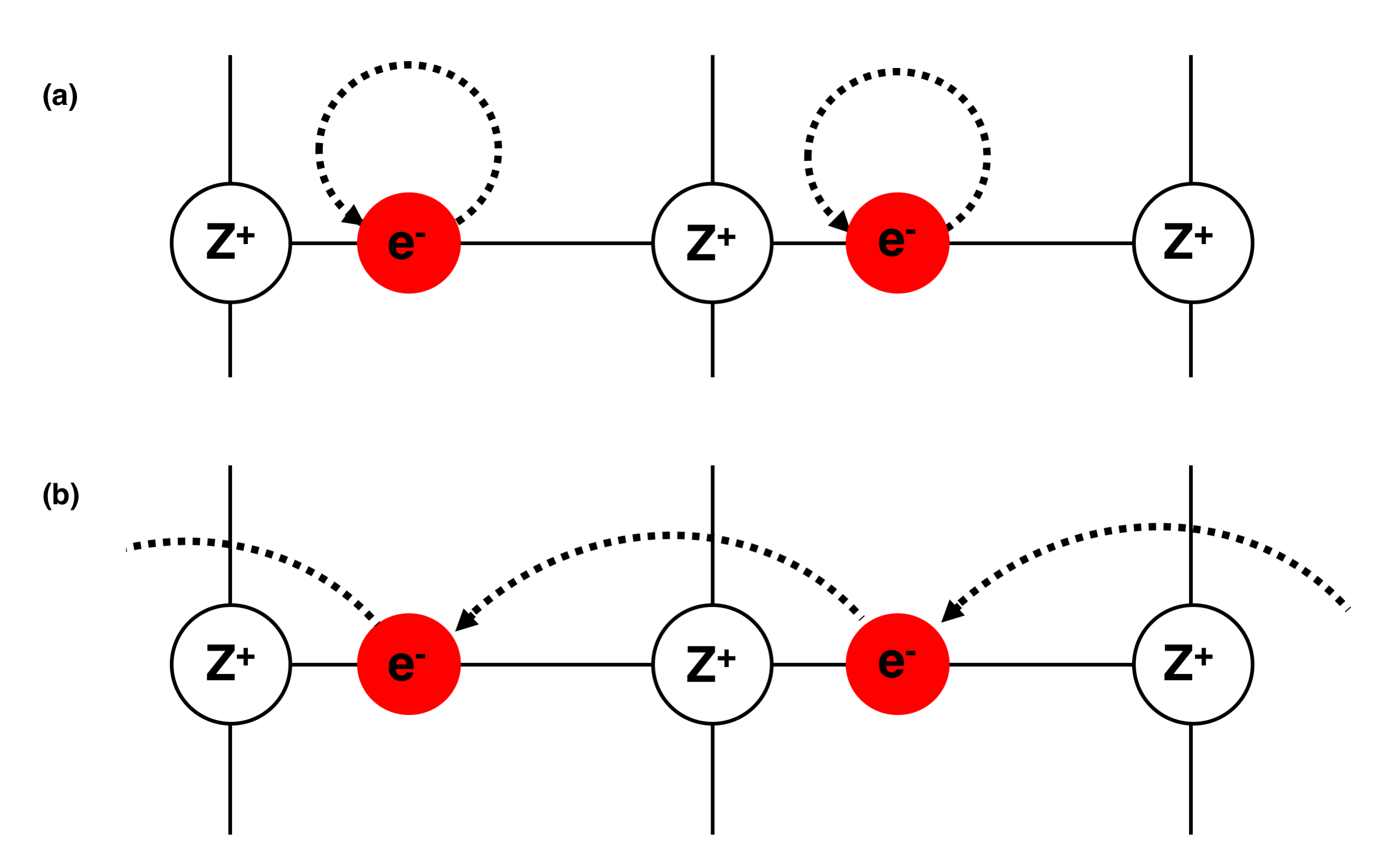}}{}
\makeatother 
\caption{{Schematic (adapted from Resta and Vanderbilt \unskip~\protect\cite{364506:8303910}) illustrating the possible trajectories for MLWF centers under an adiabatic cyclic evolution of the Hamiltonian. The MLWFCs can either \textbf{(a)} return to their initial sites or \textbf{(b)} "hop" to a new site one lattice vector away. }}
\label{f-5956}
\end{figure}
\egroup
Due to their unique connection to the Berry phase, Wannier functions are increasingly being used in the context of topological materials \unskip~\cite{364506:9893474}.  As posited by Resta and Vanderbilt, considering the behavior of Wannier centers (WCs) under a cyclic adiabatic evolution of the Hamiltonian can lead to some useful insights \unskip~\cite{364506:8303910}. At the end of such a cyclic evolution, the WCs must return to their initial positions mod the lattice constant $\mathbf R $ because the initial and final Hamiltonians are the same. As illustrated in a 1D model schematic (adapted from Resta and Vanderbilt\unskip~\cite{364506:8303910}) shown in Figure~\ref{f-5956}, one can consider two possible evolutions of the WCs that satisfy this condition.  The route involving a shift of the WCs by a lattice vector $\mathbf R $ corresponds to the quantized topological pumping phenomenon first discussed by Thouless \unskip~\cite{364506:9235357}.  In recent years, different types of Thouless pumps have been demonstrated experimentally \unskip~\cite{364506:9235339,364506:9235356}, and Wannier functions have been used as a formal means to understand the mechanisms of these dynamic topological phenomena. Most theoretical studies and descriptions of topological pumping assume complete adiabaticity of the Hamiltonian evolution. It is important to note here that topological Thouless pumping is a phenomenon that manifests from the complex phase of the wavefunctions, which means that any real MLWF method based on time-independent DFT would fail to exhibit such topological phenomenon. Our TD-MLWF/RT-TDDFT implementation, however, in principle allows us to study quantized charge transport phenomenon, including nonadiabatic effects and driving forces, from first principles.  We plan to carry out detailed studies in the future.  

Here, we demonstrate the ability for TD-MLWFs, and in particular their centers, to describe quantized charge transport in the semiconducting polymer system of trans polyacetylene. We use a quasi-monochromatic electric field pulse applied along the axis of the polymer chain as a cyclic nonadiabatic driving force for the electronic system.  Applying such a pulse to the electronic ground state of the system induces either continuous oscillations of the TD-MLWFCs or tunneling in both the +x and -x directions, depending on the magnitude of the applied electric field. In order to drive quantized charge transport in one direction, we prepare a field-polarized resonant state as the initial condition of the RT-TDDFT simulation by first adiabatically "switching on" a static electric field. This static electric field has the effect of "tilting" the periodic potential well, breaking the energy landscape symmetry in the x-direction.  Next, the field-polarized stat is subjected to a quasi-monochromatic pulse, which then drives the quantized charge transport.  In this way, we can drive quantized charge transport for a certain resonant frequency.

\bgroup
\fixFloatSize{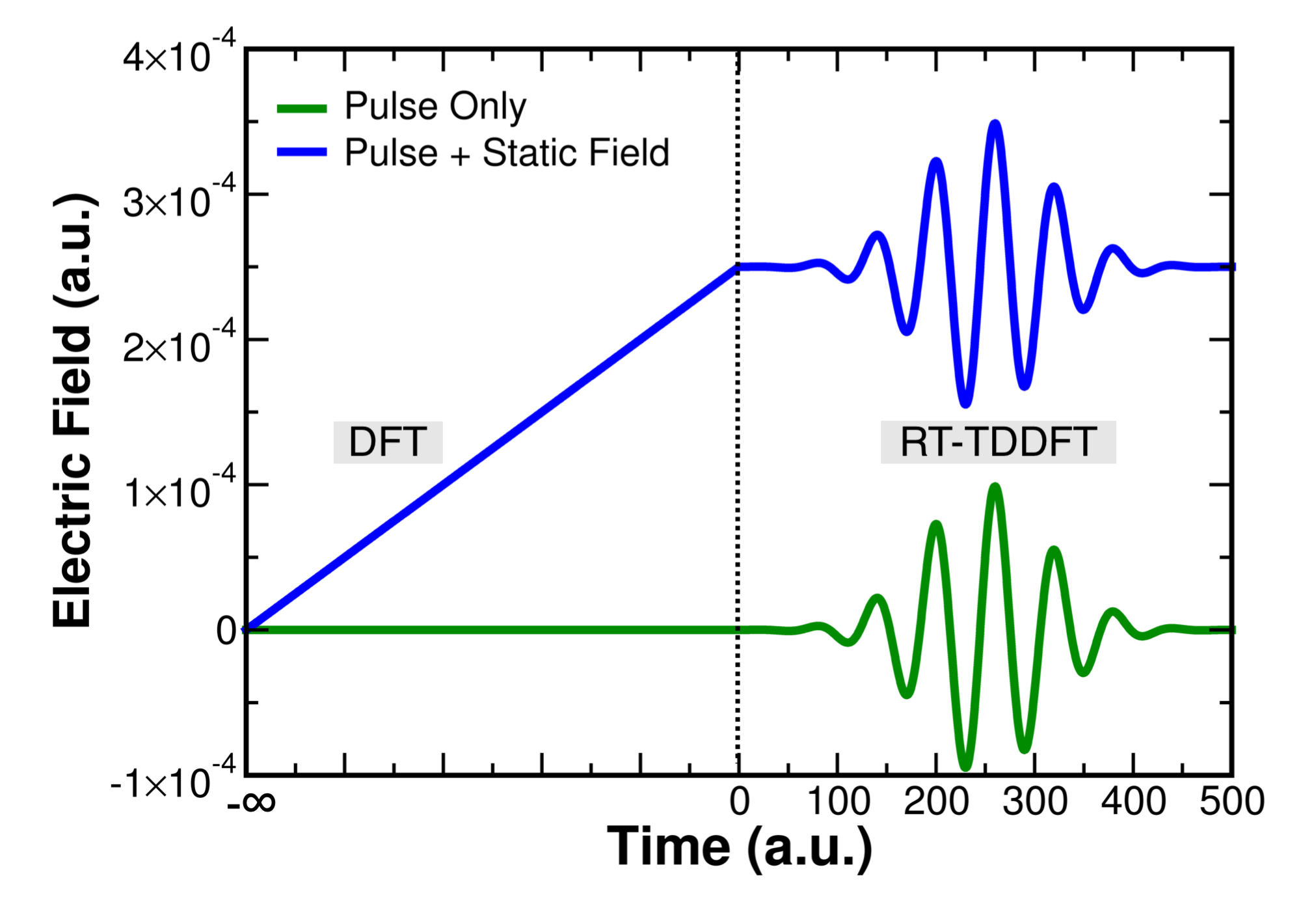}
\begin{figure}[t!]
\centering \makeatletter\IfFileExists{images/0c2589a1-48c0-4e77-be8f-c6bd12b4477c-ufig11.png}{\includegraphics{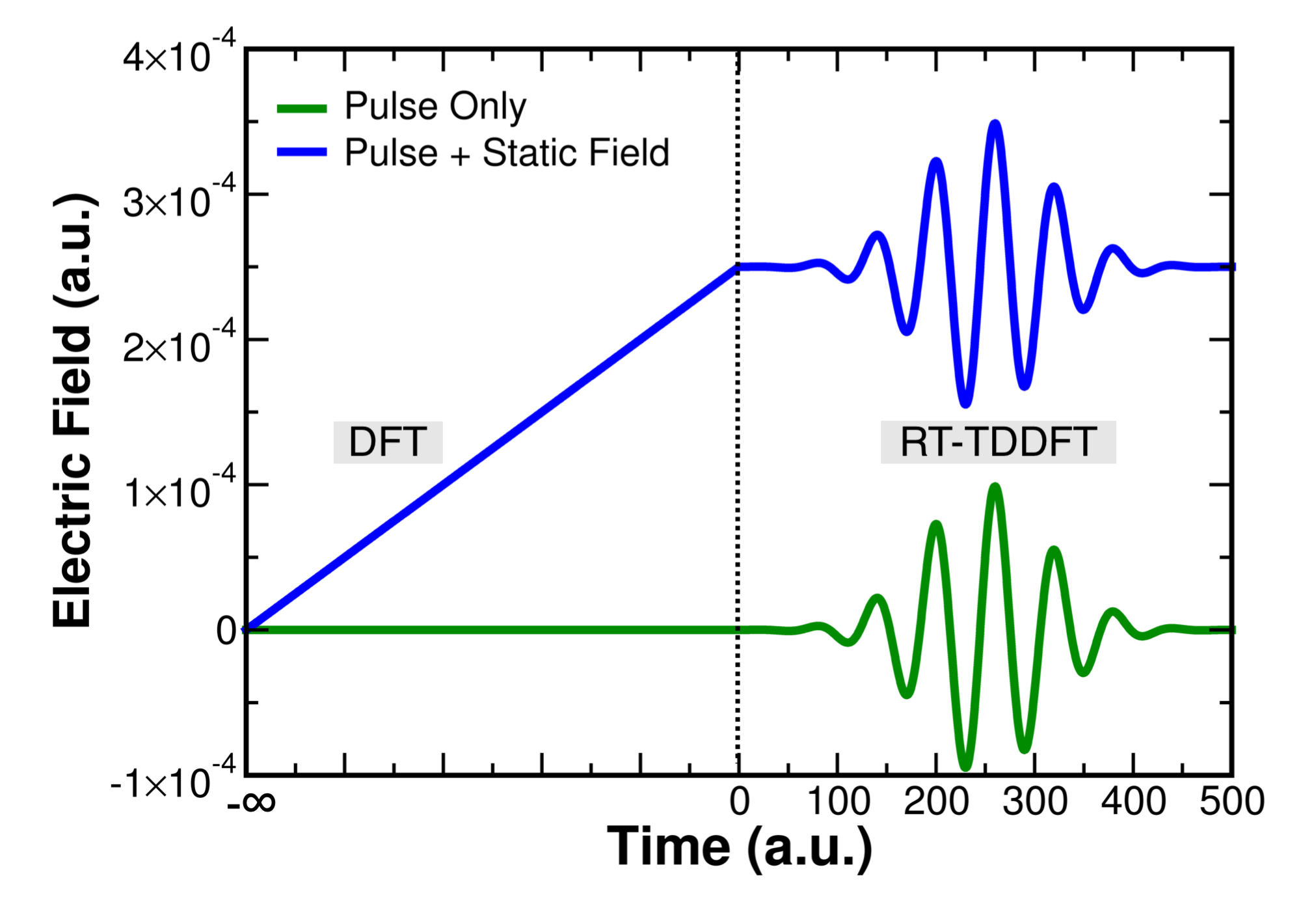}}{}
\makeatother 
\caption{{Temporal profiles of the homogeneous electric field for the two simulation cases. The region to the left of the dashed line, $t\in\left(-\infty,0\right) $ represents the adiabatic switching-on of the electric field (blue) in the time-indepedendent DFT calculation. The region to the right $t\in\left[0,500\right]\;a.u. $of the dashed line represents the beginning of the RT-TDDFT simulations with the quasi-monochromatic pulse with (blue) and without (green) the static electric field. }}
\label{f-e1d2}
\end{figure}
\egroup
The simulation details for the first-principles charge transport simulations are as follows: A 14-monomer trans-polyacetylene chain was contained in a periodic simulation cell with dimensions of (34.28 x 20.00 x 20.00 Bohr), with the polymer aligned along the x axis.  All atoms were represented by HSCV pseudopotentials. A planewave cutoff energy of 20 Ryd. was used.  The LDA XC functional was used for all calculations.  For the RT-TDDFT simulations, a 0.1 a.u. time step was used with the ETRS integrator.  Two initial electronic states for the RT-TDDFT simulations were calculated: 1) the electronic ground state calculated via DFT, and 2) the field-polarized state calculated adiabatically via DFT in the presence a static electric field in the +x direction with a magnitude of $2.5\;\times\;10^{-3} $ a.u.  For the RT-TDDFT simulations, these systems are then subjected to a quasi-monochromatic electric field pulse with a frequency of 2.8 eV (calculated resonant frequency of double-bond TD-MLWFs), a maximum field-strength of $1.0\;\times\;10^{-3} $ a.u., which is enveloped in a gaussian with a full-width at half-maximum of 1.0 eV.  In Figure~\ref{f-e1d2}, the temporal profiles of the electric fields are shown, with the adiabatic (DFT) and nonadiabatic (RT-TDDFT) portions of the simulations delineated.

\bgroup
\fixFloatSize{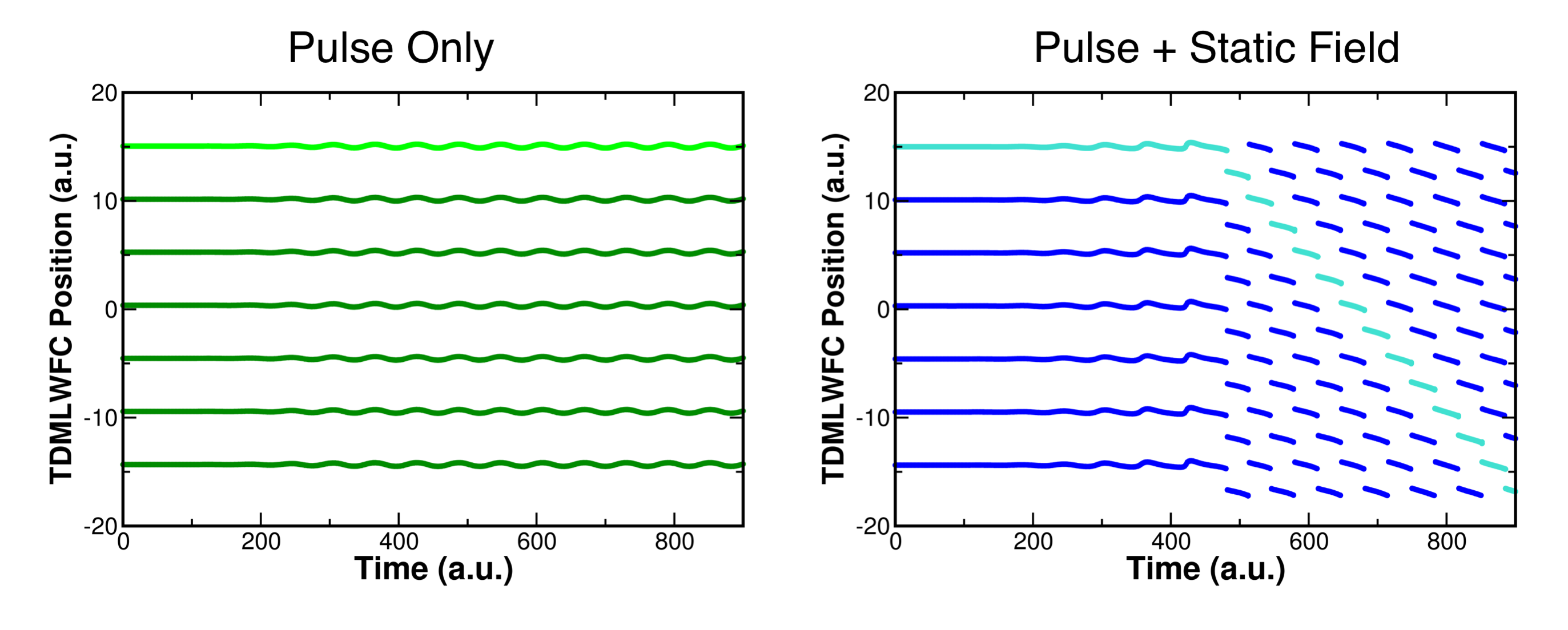}
\begin{figure*}[t!]
\centering \makeatletter\IfFileExists{images/9686459f-0cee-4548-b7eb-f4edfdef99ab-ufig12.png}{\includegraphics[width=.74\linewidth]{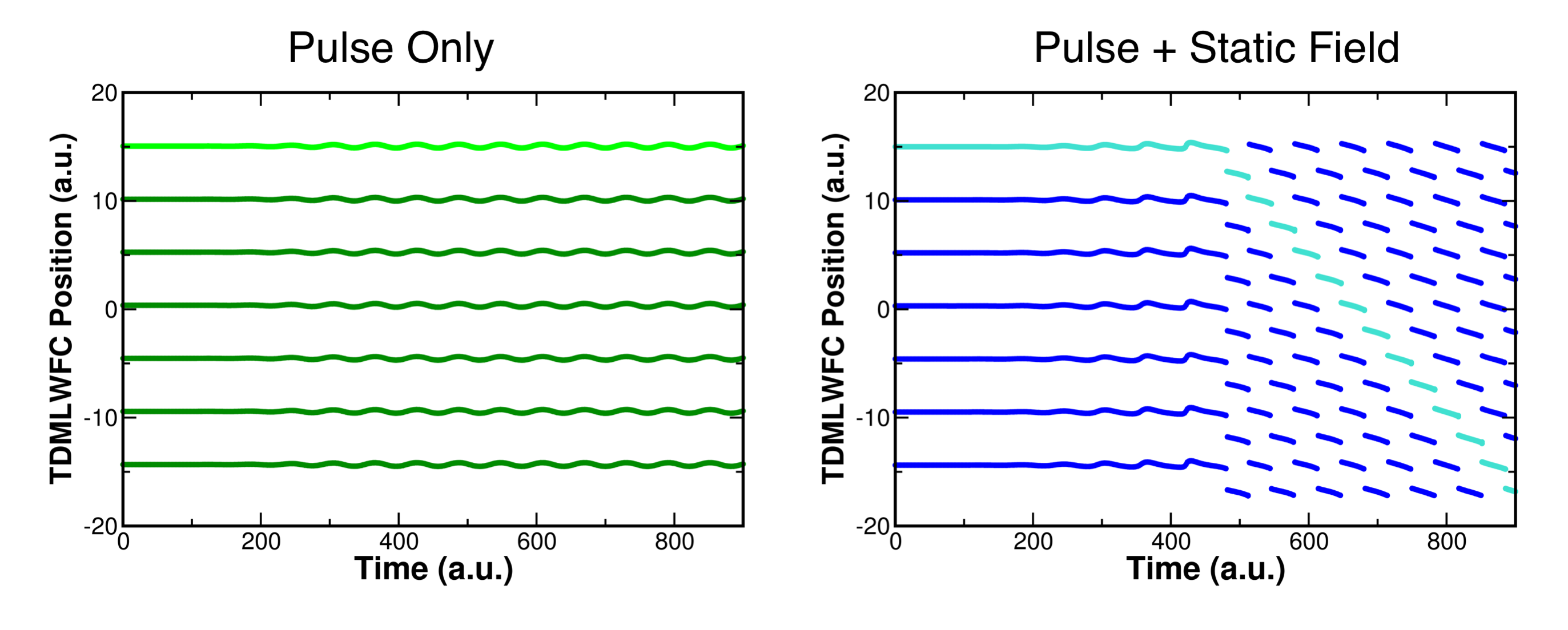}}{}
\makeatother 
\caption{{Positions of the double-bond TD-MLWFCs on the polyacetylene chain (commensurate with the x-axis) throughout the RT-TDDFT simulations for the "pulse only" (left) and the "pulse + static field" (right) cases. The positions of the TD-MLWFCs are plotted as a point at each step in the RT-TDDFT simulation to show the presence of the jump discontinuities (right). \textbf{\space }}}
\label{f-773b}
\end{figure*}
\egroup

\bgroup
\fixFloatSize{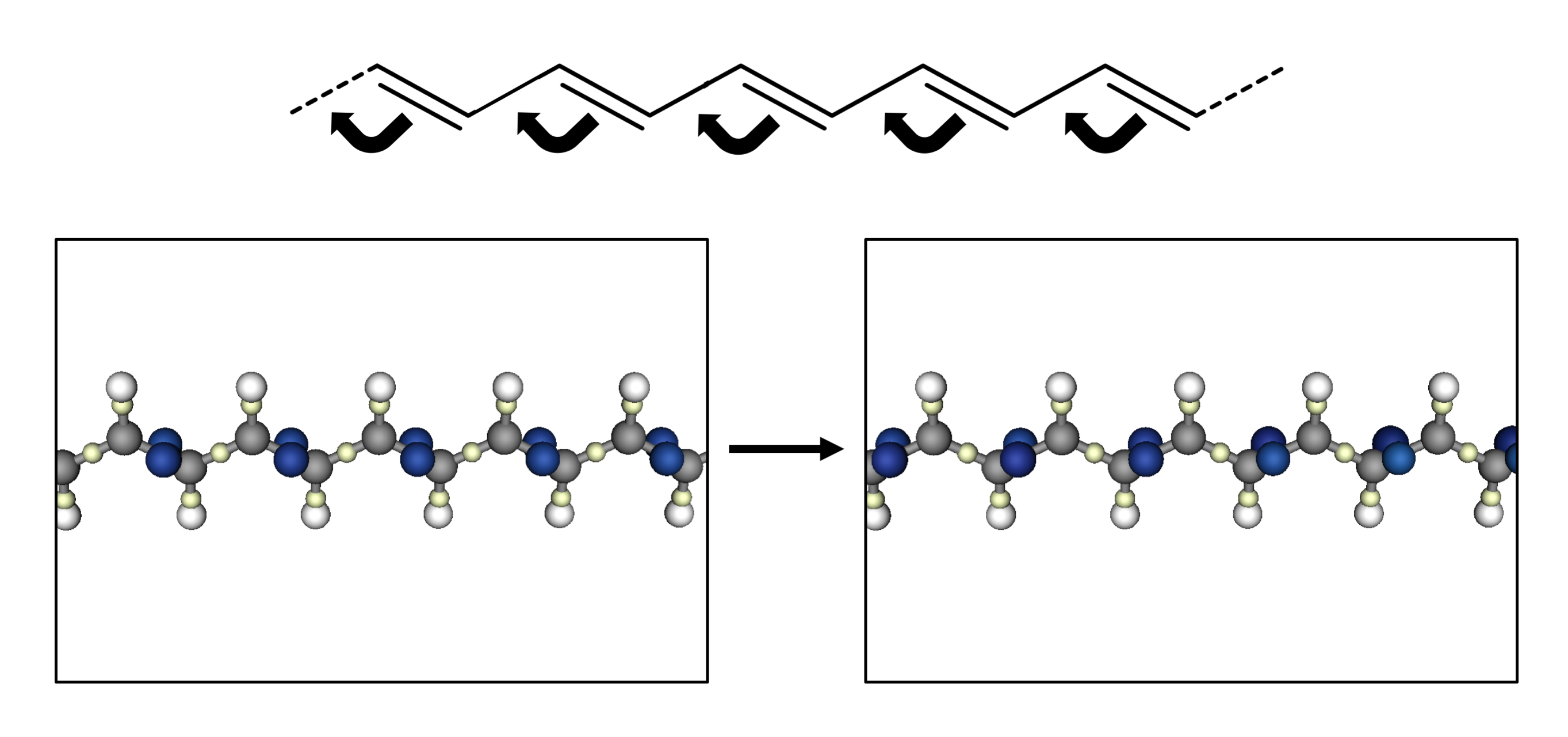}
\begin{figure}[b!]
\centering \makeatletter\IfFileExists{images/2d59d9ab-756d-4c50-b7e8-16abc0055eaa-ufig13.png}{\includegraphics{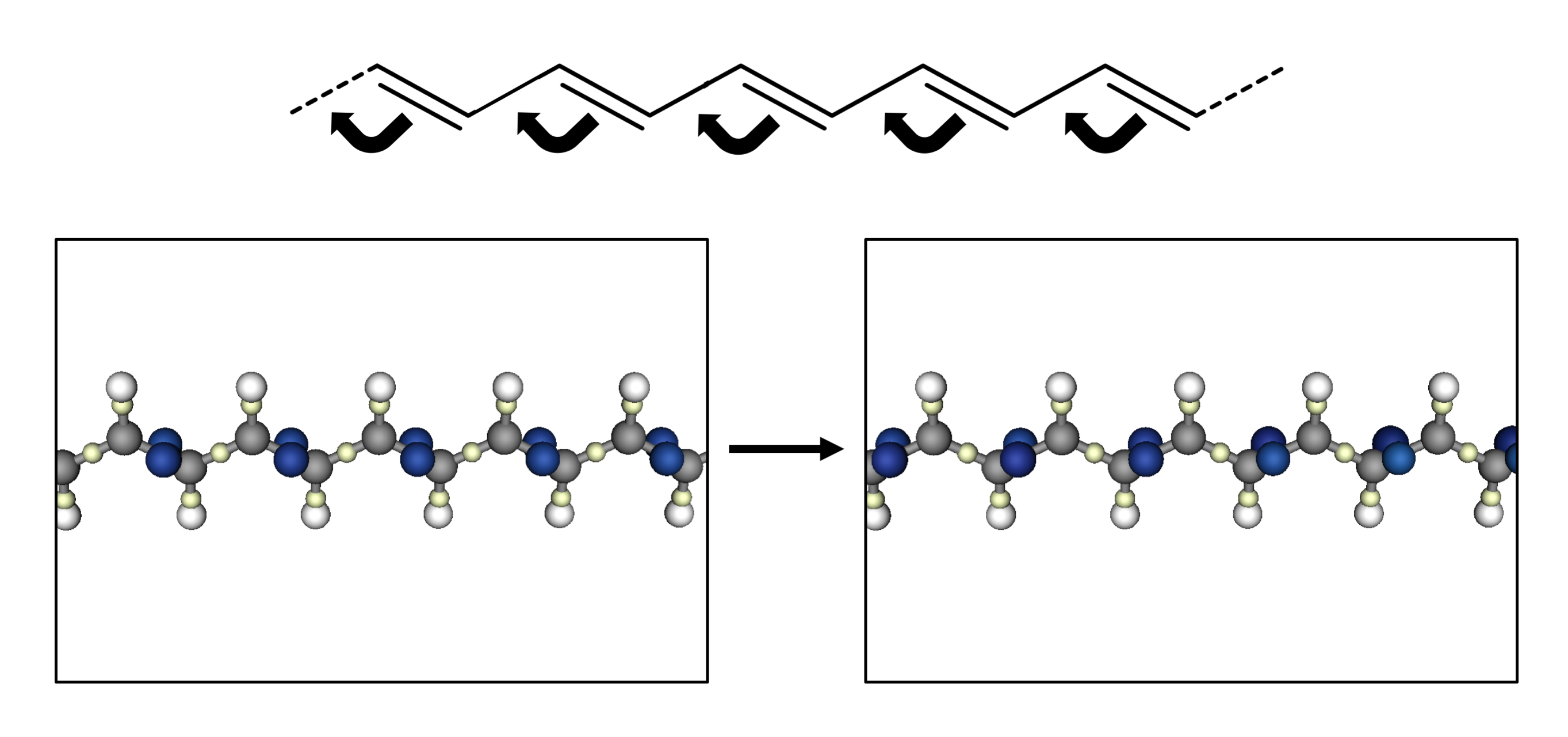}}{}
\makeatother 
\caption{{(Upper) Skeletal chemical structure for the trans polyacetylene chain illustrating the double-bond TD-MLWFCs hopping occurring in the RT-TDDFT simulation. (Lower) Two consecutive RT-TDDFT simulation snapshots showing the carbon and hydrogen atoms (grey and white) and the TD-MLWFCs (yellow and dark blue) as spheres. Here, the TD-MLWFC spheres have been scaled and colored proportionally to the spreads of the associated TD-MLWFs. }}
\label{f-70b6}
\end{figure}
\egroup
The spatial x-axis coordinates of the double-bond TD-MLWFCs with respect to time are shown in Figure~\ref{f-773b} for both simulation cases.  Although the electrons, and consequently the TD-MLWFs, are all in principle indistinguishable, in Figure~\ref{f-773b} we have highlighted some of the plotted data as a visual guide to understand the observed phenomena.  In the pulse-only case (Figure~\ref{f-773b} left), the dynamics are fairly simple, with the quasi-monochromatic pulse exciting a resonance in the double-bond TD-MLWFs that causes repetitive Rabi oscillations, but no net charge transport in one direction. With the presence of the static electric field in addition to the quasi-monochromatic pulse (Figure~\ref{f-773b} right), however, the dynamics are clearly different. At a point in time near the end of the pulse ({\texttildeapprox}475 a.u.), there is a sudden jump in the TD-MLWFC positions.  This discontinuity, an average of 1.8 a.u. displacement in the -x direction, corresponds to double-bond TD-MLWFs tunneling from one site on the polyacetylene chain to another. Figure~\ref{f-70b6}  illustrates this pictorially through the perspective of the chemical structure and by showing snapshots of the polyacetylene TD-MLWFs from one RT-TDDFT time step to the next, only 0.1 a.u. later.  For both simulations all nuclei were held in fixed positions. Thus, even after the applied electric field is turned off at t = 500 a.u., the electronic system continues to propagate freely and cyclically as in a Rabi oscillation.  Consequently, we observe that the quantized charge transport, initially driven by the pulse, continues repetitively, regularly, on average every 67.9 a.u. of time and with -1.81 a.u. displacement of each TD-MLWFC.  The discretized tunneling phenomenon is also reflected in the TD-MLWF spread.  Figure~\ref{f-35eb} shows the total change in spread of the TD-MLWFs throughout the RT-TDDFT simulations.  For the pulse-only case, the dynamics are as expected, with the pulse having the effect of cyclic oscillations in the spread (i.e., oscillatory localization and delocalization of the TD-MLWFs).  However, in the pulse + static field case, we observe again jump discontinuities simultaneous with the TD-MLWFC position jumps.  As the TD-MLWFs come close to tunneling, the spread increases rapidly, and then suddenly decrease after the tunneling occurred and the TD-MLWFs return to being  localized between a different pair of carbon atoms. 
\bgroup
\fixFloatSize{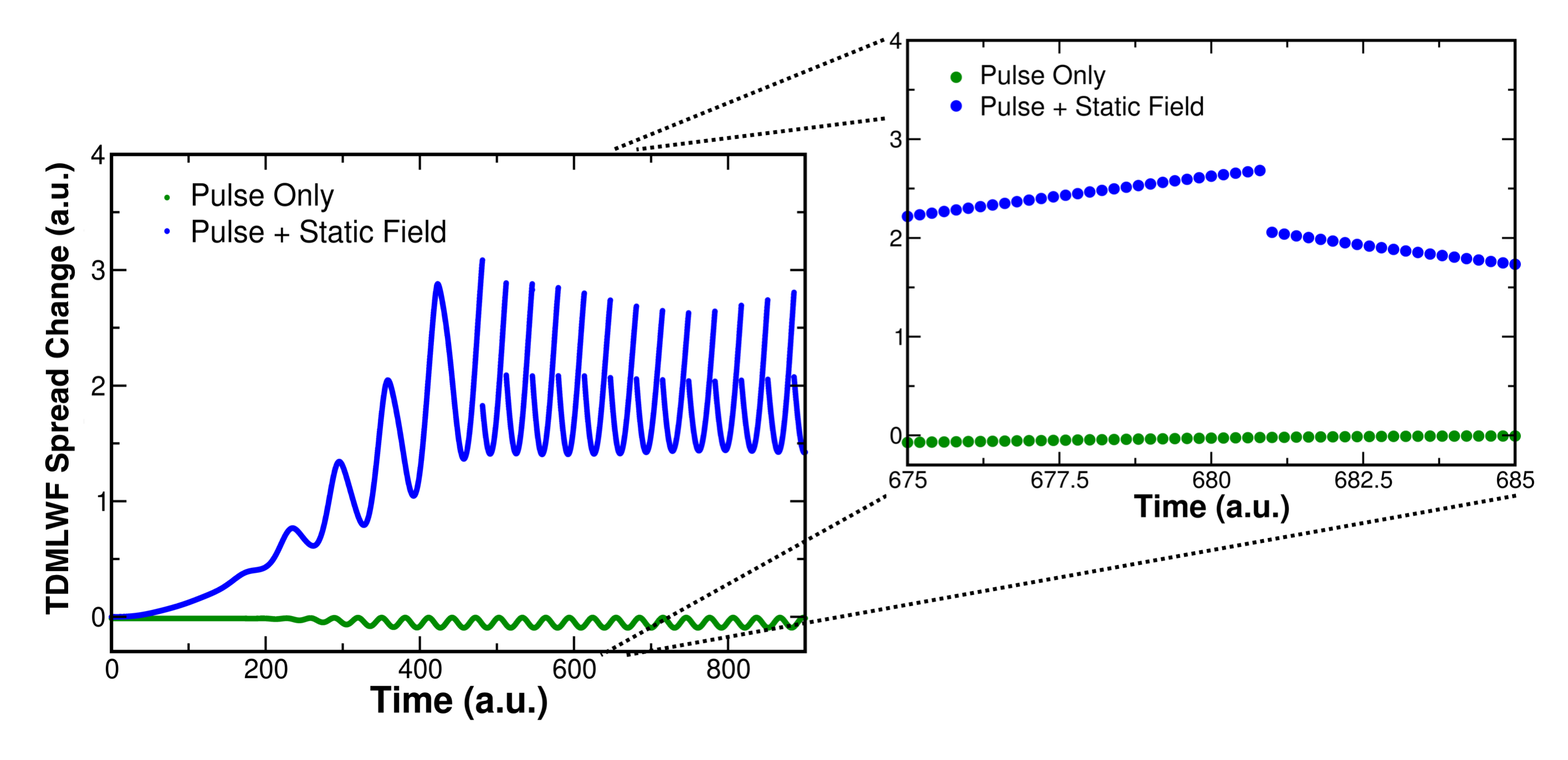}
\begin{figure*}[t!]
\centering \makeatletter\IfFileExists{images/3efda505-bcf0-4c10-b3a9-b02c2873d734-ufig14.png}{\includegraphics[width=.73\linewidth]{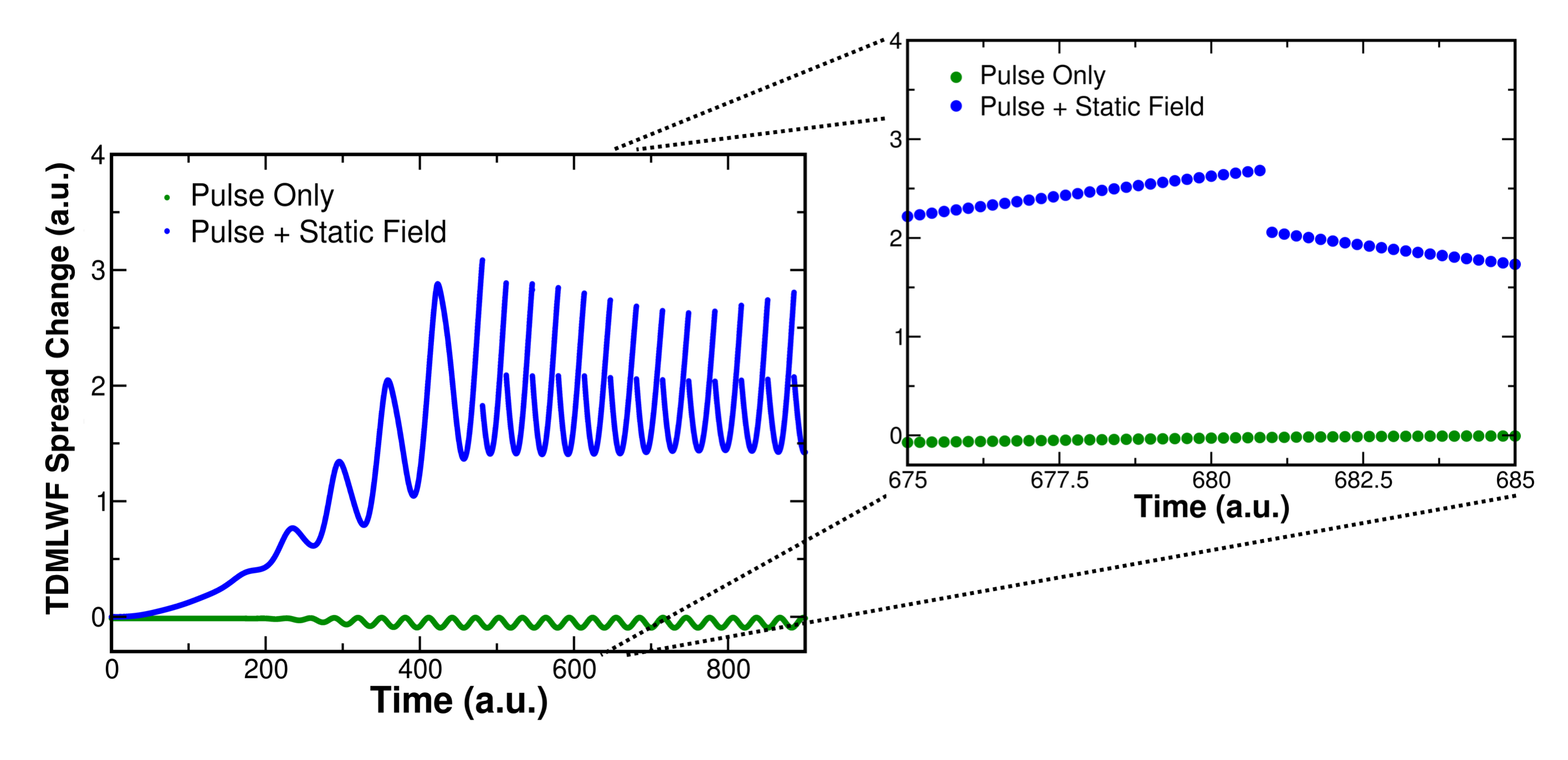}}{}
\makeatother 
\caption{{Total change in the TD-MLWF spread throughout the RT-TDDFT simulations for the "pulse only" case (green dots) and the "pulse + static field" case (blue dots). The inset zooms in on a small time region to illustrate the sudden decrease in spread at the time of TD-MLWF tunneling. }}
\label{f-35eb}
\end{figure*}
\egroup

\subsection{Electronic Stopping Excitations}
\bgroup
\fixFloatSize{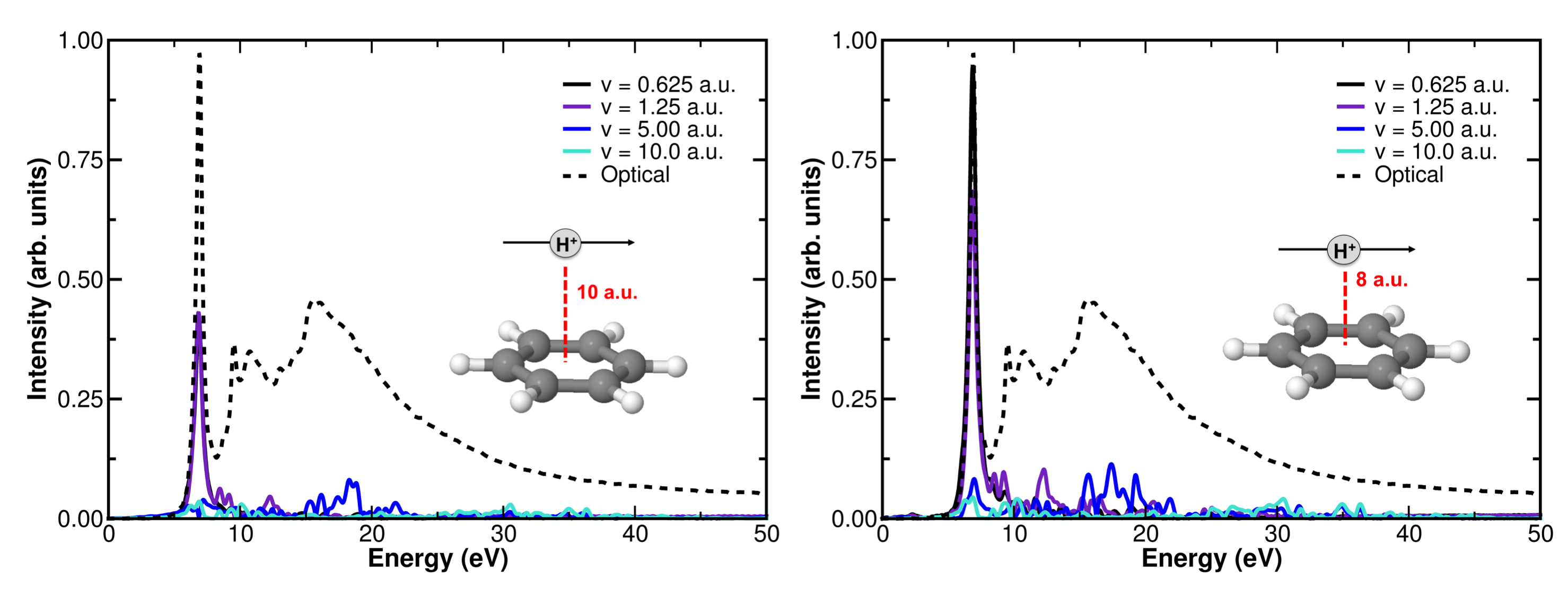}
\begin{figure*}[t!]
\centering \makeatletter\IfFileExists{images/cee49b28-4f74-4a52-bcc9-0f94b6ddb287-ufig15.png}{\includegraphics{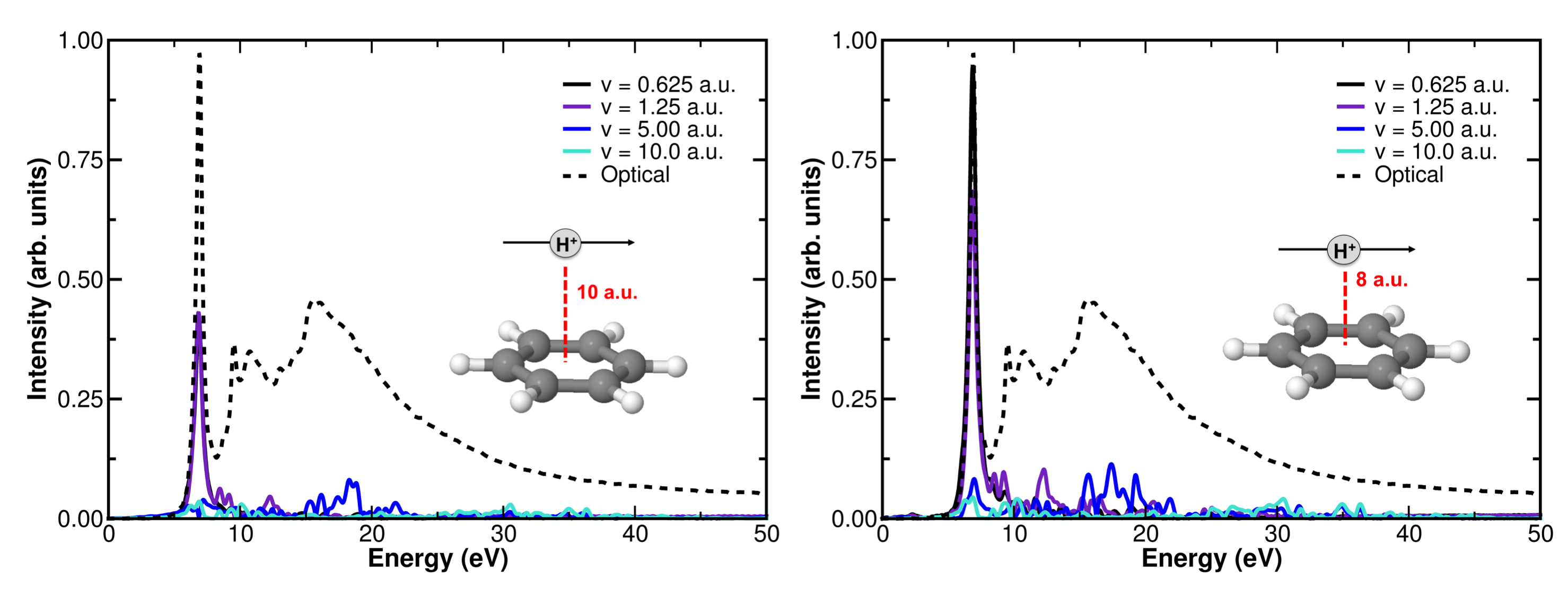}}{}
\makeatother 
\caption{{"Electronic stopping spectra" for a proton impinging on a gas phase benzene molecule with an impact parameter of 10 a.u. (left) and an impact parameter of 8 a.u. (right). Four velocities (0.625, 1.25, 5.00, and 10.0 a.u.) were simulated and four spectra were calculated for each case. The optical spectrum (dashed line) is shown for comparison. }}
\label{f-e5c5}
\end{figure*}
\egroup
One physical process that RT-TDDFT has been quite successful in simulating is electronic stopping. Electronic stopping is the phenomenon that occurs when fast-moving charged particles ({\textgreater} {\texttildeapprox}10 keV) transfer energy to the surrounding electronic system through excitation and ionization. Electronic stopping power is defined as the rate of energy transfer from projectile ion to electrons, is a key quantity in describing this process. In recent years, RT-TDDFT simulations have been quite successful in predicting the electronic stopping power of a range of matter and for a variety of projectile ions \unskip~\cite{364506:8217172,364506:8217218,364506:8217261,364506:8217219,364506:8094190}. However, there are more scientific insights to be gained from understanding these ion irradiation excitation dynamics on an atomistic level. Induced density analysis \unskip~\cite{364506:8217261} and single-particle projection methods \unskip~\cite{364506:8297331} have been previously used as tools to understand the molecular level details of the stopping process. However,  both of these approaches have limitations. Analysis in terms of the time-dependent induced density can be difficult to interpret, and quantitative comparisons usually require the use of a charge-partitioning scheme, which relies on a reasonable, but still rather arbitrary, choice. Projection methods, on the other hand, also have their limitations.  In particular, with projections onto the equilibrium (i.e. ground state) eigenstates, it is difficult to include enough virtual/unoccupied eigenstates in the projection calculation to account for excited electrons in ionization, which is rather common in electronic stopping process. Thus, in practice, projection methods are only most useful in analyzing excited hole distributions. With these challenges in mind, TD-MLWF propagation provides a new, useful tool.  By construction, TD-MLWFs can give molecular level details of the excited electrons in the electronic stopping process without the need to resort to charge density partitioning scheme. Additionally, a key challenge in studies of electronic stopping is to examine differences in the electron excitation induced by photon and ion radiation. The TD-MLWF approach, with its flexibility in simulations of both types of radiation, can aid in this challenge.  

In order to demonstrate some of the capabilities of TD-MLWFs for descriptions of electronic stopping processes, we have performed several RT-TDDFT simulations of proton irradiation of an isolated benzene molecule.  The simulations were performed with a 0.1 a.u. time step in the ETRS propagator, a 30 Rydberg planewave kinetic cutoff energy, the LDA XC functional, and HSCV pseudopotentials. The setup was as follows: a proton with constant velocity in the +x direction was moved through a 30 x 30 x 50 a.u. simulation cell containing a benzene molecule oriented in the xy plane.  Four proton velocities were simulated (v = 0.625, 1.25, 5.00, 10.0 a.u.) at two different impact parameters relative to the plane of the benzene molecule. Once the proton has traversed the length of the simulation cell (50 a.u.), it is removed, at which point the RT-TDDFT simulation and TD-MLWF propagation is continued for 500 a.u. The electronic dipole moment of the excited benzene molecule, calculated via the TD-MLWFCs, then continues to oscillate. Taking the absolute value of the Fourier transform of the electronic dipole moment presents a spectrum of the excitations caused in the proton electronic stopping process.  Figure~\ref{f-e5c5} shows the resultant "electronic stopping spectra", alongside the calculated optical spectrum.  For the low-velocity cases (0.625 and 1.25 a.u.), the spectra primarily comprise a single well-defined peak at {\texttildeapprox}6.9 eV, which aligns with the lowest energy peak in the optical spectrum, indicating that at lower velocities, the proton primarily excites the mode that corresponds to the first optical excitation. Approaching higher proton velocities (5.00 and 10.0 a.u.), the peak at {\texttildeapprox}6.9 eV diminishes, and we emergence of peaks in higher energy ranges (15 eV to 40 eV).  This result is consistent with previous RT-TDDFT studies of electronic stopping which have shown that more ionization occurs as the ion velocity increases. We also observed an impact parameter dependence in the electronic stopping spectra (see Figure~\ref{f-e5c5}, left panel compared to right panel). Not surprisingly, there is an overall trend of increasing spectrum intensity with decreasing impact parameter for all velocities. However, the low-velocity spectra (0.625 a.u. and 1.25 a.u.) show large increases in intensity (by a factor or {\texttildeapprox}2), whereas the high-velocity spectra show much smaller increases. 
    
\section{Conclusions}
Real-time, time-dependent density functional theory (RT-TDDFT) has gained popularity as a first-principles approach to study a variety of excited-state phenomena such as optical excitations and electronic stopping. Within RT-TDDFT simulations, the gauge freedom of the time-dependent electronic orbitals can be exploited for numerical and scientific convenience while the unitary transformation does not alter physical properties calculated from the quantum dynamics of electrons. Exploiting this gauge freedom, we demonstrated the propagation of maximally localized Wannier functions (MLWFs) in RT-TDDFT within our plane-wave pseudopotential RT-TDDFT branch of the QB@LL code. 

In recent years, the gauge freedom in RT-TDDFT has been exploited in various ways including using the so-called parallel transport gauge for numerical efficiency \unskip~\cite{364506:8321071}. The time-dependent MLWF (TD-MLWF) formalism detailed in this work represents another such gauge representation that is practical for a wide range of RT-TDDFT simulations.  Due to the MLWF connection to the Berry Phase, the dynamic polarization of both isolated molecular systems and condensed matter systems in periodic boundary conditions are well-defined, allowing for a scalar potential representation of homogeneous electric fields.  This length gauge formulation circumvents the commonly-used velocity gauge \unskip~\cite{364506:8092425} formulation (within the electrodynamics) in which the electric field needs to be represented as a magnetic flux when periodic boundary conditions are used. As a demonstration, we have performed RT-TDDFT simulations of the linear response of both molecular and crystalline systems in order to calculate the resultant absorption spectra. For the case of benzene, we observe good agreement between the RT-TDDFT calculated and experimental absorption spectra.  As another molecular case, we used the TD-MLWF method to calculate the absorption spectrum of an isolated (i.e., gas phase) water molecule.  In this case, we observe excellent qualitative agreement with the experimental optical absorption spectrum covering a wide range of excitation energies (0 - 150 eV). Notably, the simulations do not give rise to any spurious peaks in the high energy region, a boon that is likely a result of the planewave pseudopotential basis upon which the RT-TDDFT code is implemented. Additionally, we have demonstrated the utility of TD-MLWFs for analyzing intermolecular and intramolecular excitation dynamics.  In the case of the isolated molecules (benzene and water), we demonstrated the decomposition of the optical absorption spectra in terms of different chemical moieties, such as lone-pairs oxygen-hydrogen bonds, carbon-carbon bonds, etc. We showed how this TD-MLWF decomposition scheme can be used also to examine the absorption spectrum of a benzene molecule solvated in liquid water, demonstrating the possibilities for investigating solvatochromic effects. As a capability demonstration of the length-gauge formulation for representing homogeneous electric field in periodic boundary conditions, we calculated the dielectric function of crystalline silicon, a system which has been well-studied with other TDDFT codes. Our spectrum shows a good agreement with both LR-TDDFT results \unskip~\cite{364506:8092474} and RT-TDDFT results from a variety of codes in atomic orbital \unskip~\cite{364506:8092425} and real-space grid \unskip~\cite{364506:8298307} implementations.  

One application of RT-TDDFT that was not discussed in this work is the computation of core-level optical spectra, such as X-ray absorption spectra (XAS). While we have recently studied core electron excitation dynamics in electronic stopping successfully \unskip~\cite{364506:10965073}, there remain significant challenges in the computation of XAS with RT-TDDFT for a number of reasons. In the context of atomic-orbital based RT-TDDFT simulations, such calculations have been carried out in recent years \unskip~\cite{364506:8217170,364506:8351790,364506:8217423,364506:10964664}.  However, with planewave pseudopotential (PW-PP) approaches, accurate representation of core electron wavefunctions requires an extremely high planewave cutoff (i.e., PW basis set expansion) which makes the calculation of XAS particularly computationally difficult in practice. And in general, for any underlying basis set of choice, there remain shortcoming with regards to the exchange-correlation approximation, with popular hybrid functionals even failing to accurately predict XAS excitation energies \unskip~\cite{364506:10964664}.  Although the TD-MLWF approach can be used also for modeling of XAS using RT-TDDFT, it is a topic that requires a thorough investigation in a future work.  

We presented a few examples of the great utility of the TD-MLWF gauge representation beyond the linear response regime (e.g. optical absorption spectrum). Using the ability to simulate both static and time-dependent electric fields as scalar potentials in the length gauge, we demonstrated the simulation of quantized charge transport in polyacetylene. Although it is a relative simple test case, similar simulations could be used in the future to carry first-principles investigations into dynamic topological transport phenomena, beyond the adiabatic dynamic description of a topological Thouless pump \unskip~\cite{364506:9235357,364506:9235358}. Finally, we presented TD-MLWF RT-TDDFT simulations of the electronic stopping of a proton impinging on a benzene molecule. Although it is often difficult to make direct comparisons between the excitation behavior of systems under photon and ion irradiation, we have proposed a type of analysis through the calculation of "electronic stopping spectrum" based on the dynamic polarization in the same way optical absorption spectrum is calculated from the dynamics polarization. Using the electronic stopping spectrum, one can identify what parts of the optical absorption spectrum are excited in electronic stopping process as a function of the projectile proton velocity.

There are many additional applications of RT-TDDFT where TD-MLWFs could provide advantages over conventional TDKS propagation. In the context of ground-state DFT, the spatial locality of MLWFs have been used to greatly accelerate calculations with hybrid exchange-correlation (XC)  functionals \unskip~\cite{364506:8842590} and first principles molecular dynamics simulations \unskip~\cite{364506:8842632}. We have demonstrated that, for large systems, the TD-MLWF implementation scales well over tens of thousands of cores without incurring prohibitive computational costs (i.e. only two times more expensive). Thus, the TD-MLWF approach makes it possible to also accelerate hybrid XC calculations in the context of RT-TDDFT for studying large complex condensed matter systems.
\begin{acknowledgments}The authors would like to thank Erik W. Draeger for helpful discussions. This work is supported by the National Science Foundation under Grants No. CHE-1565714, No. DGE-1144081, and No. OAC-17402204.  An award of computer time was provided by the Innovative and Novel Computational Impact on Theory and Experiment (INCITE) program. This research used resources of the Argonne Leadership Computing Facility, which is a DOE Office of Science User Facility supported under Contract DE-AC02{\textendash} 06CH11357
\end{acknowledgments}
    
\linespread{1.5} 

\bibliographystyle{aipnum4-1}

\bibliography{\jobname}
\end{document}